\PassOptionsToPackage{unicode}{hyperref}
\PassOptionsToPackage{hyphens}{url}
\documentclass[11pt,a4paper]{article}

\usepackage[textwidth=5.5in, textheight=9in, headheight=14pt, centering]{geometry}

\usepackage{amsmath,amssymb}
\usepackage{iftex}
\ifPDFTeX
  \usepackage[T1]{fontenc}
  \usepackage[utf8]{inputenc}
  \usepackage{textcomp}
  \usepackage{mathpazo}  
\else 
  \usepackage{unicode-math}
  \defaultfontfeatures{Scale=MatchLowercase}
  \defaultfontfeatures[\rmfamily]{Ligatures=TeX,Scale=1}
\fi
\ifPDFTeX\else
\fi

\ifPDFTeX
  \usepackage[protrusion=true,expansion=true,tracking=true,kerning=true,spacing=true]{microtype}
\else
  \usepackage[protrusion=true]{microtype}
\fi
\UseMicrotypeSet[protrusion]{basicmath}

\usepackage[table,dvipsnames]{xcolor}
\definecolor{okblue}{RGB}{0,114,178}
\definecolor{okorange}{RGB}{230,159,0}
\definecolor{okgreen}{RGB}{0,158,115}
\definecolor{okred}{RGB}{213,94,0}
\definecolor{okpurple}{RGB}{204,121,167}
\definecolor{okyellow}{RGB}{240,228,66}
\definecolor{oksky}{RGB}{86,180,233}
\definecolor{darkblue}{RGB}{0,70,130}
\definecolor{linkblue}{RGB}{0,90,160}
\definecolor{rowshade}{gray}{0.93}

\usepackage{parskip}

\usepackage{longtable,booktabs,array}
\usepackage{calc}
\usepackage{etoolbox}
\usepackage{colortbl}
\usepackage{threeparttable}
\makeatletter
\patchcmd\longtable{\par}{\if@noskipsec\mbox{}\fi\par}{}{}
\makeatother
\IfFileExists{footnotehyper.sty}{\usepackage{footnotehyper}}{\usepackage{footnote}}
\makesavenoteenv{longtable}
\renewcommand{\arraystretch}{1.2}

\usepackage{graphicx}
\usepackage[
  font=scriptsize,
  labelfont=bf,
  labelsep=period,
  skip=6pt,
  margin=0.5cm
]{caption}
\makeatletter
\newsavebox\pandoc@box
\newcommand*\pandocbounded[1]{%
  \sbox\pandoc@box{#1}%
  \Gscale@div\@tempa{\textheight}{\dimexpr\ht\pandoc@box+\dp\pandoc@box\relax}%
  \Gscale@div\@tempb{\linewidth}{\wd\pandoc@box}%
  \ifdim\@tempb\p@<\@tempa\p@\let\@tempa\@tempb\fi%
  \ifdim\@tempa\p@<\p@\scalebox{\@tempa}{\usebox\pandoc@box}%
  \else\usebox{\pandoc@box}%
  \fi%
}
\def\fps@figure{htbp}
\makeatother

\usepackage{titlesec}
\titleformat{\section}
  {\Large\sffamily\bfseries}{\thesection}{1em}{}
\titleformat{\subsection}
  {\large\sffamily\bfseries}{\thesubsection}{1em}{}
\titleformat{\subsubsection}
  {\normalsize\sffamily\bfseries}{\thesubsubsection}{1em}{}
\titlespacing*{\section}{0pt}{24pt plus 4pt minus 2pt}{12pt plus 2pt minus 2pt}
\titlespacing*{\subsection}{0pt}{18pt plus 3pt minus 1pt}{8pt plus 2pt minus 1pt}
\titlespacing*{\subsubsection}{0pt}{12pt plus 2pt minus 1pt}{6pt plus 1pt minus 1pt}

\usepackage{fancyhdr}
\fancypagestyle{plain}{%
  \fancyhf{}%
  \fancyfoot[C]{\small\thepage}%
}

\usepackage{enumitem}
\setlist{nosep, leftmargin=*, font=\small}
\setlist[itemize]{label=\textbullet}
\setlist[enumerate]{label=\arabic*.}

\providecommand{\tightlist}{%
  \setlength{\itemsep}{0pt}\setlength{\parskip}{0pt}}
\usepackage{bookmark}
\IfFileExists{xurl.sty}{\usepackage{xurl}}{}
\hypersetup{
  colorlinks=true,
  linkcolor=darkblue,
  citecolor=okblue,
  urlcolor=linkblue,
  pdfcreator={LaTeX via pandoc}}
\usepackage{pdflscape}
\usepackage[nameinlink,noabbrev]{cleveref}

\IfFileExists{upquote.sty}{\usepackage{upquote}}{}

\author{Maximilian Schons, MD\textsuperscript{1} \and Red Bermejo\textsuperscript{4} \and Florian Aldehoff-Zeidler\textsuperscript{2} \and Niccolò Zanichelli\textsuperscript{4} \and Oliver Evans\textsuperscript{4} \and Gavin Leech, PhD\textsuperscript{3} \and Samuel Härgestam\textsuperscript{4}}
\date{}

\begin{document}

\title{How much technical talent is there? A systematic estimate of the ML research pool among 3 million IT consultancy employees}
\maketitle
\begin{center}
{\small\itshape \textsuperscript{1}MxSchons GmbH \quad \textsuperscript{2}three backticks (0x60.net) \quad \textsuperscript{3}Arb Research \quad \textsuperscript{4}Independent}
\end{center}


\section{Abstract}\label{h.mwqxvzn3cald}

\subsection{Why was this study done?}\label{h.memzwv7csb2}

To determine whether there is latent capacity of capable technical
machine learning research talent in the IT Consultancy sector. This
talent pool could advance AI risk mitigation and alignment research
agendas.

\subsection{How was the study conducted?}\label{h.ld60q4kcq1u1}

We systematically searched the internet, global business databases, and
conference/paper affiliations for ML consulting firms. Employee LinkedIn
resumes were then scored by keyword filters and large-language-model
(LLM) classifiers; these signals were combined in a bootstrap probit
model to estimate technical ML Research Talent per firm. A subset of
companies also completed a 3-day ML research \& engineering work trial.

\subsection{Results}\label{h.snhs9eyfaxcm}

We screened 2\,121 organizations and found 403 to offer broader ML
related consulting services. 3\,269\,000 employees were associated with
this sample. The distribution was 284 (70.5\%) small (\textless{} 100
employees), 76 (18.9\%) medium (100--999 employees), 23 (5.7\%) large
(1\,000--9\,999 employees), and 20 (5.0\%) giant ($\geq$10\,000 employees)
companies. The 50th percentile aggregate estimate of highly technical
ML Research Talent across these organizations was 1\,121 (80\% CI: 252--3\,165).

For our work trial 97 companies were approached, 20 applied, 8 were
invited to participate, and 5 of 8 (63\%) received at least a conditional
recommendation for technical AI safety work. No AI model was able to
pass the work trial successfully.

\subsection{Limitations and future research}\label{h.wort6x2n8vb2}

Some companies, particularly in certain geographic regions, might be
poorly represented on LinkedIn. More generally, resumes remain an
inherently noisy signal of competence and our definition of technical ML
research talent might have excluded some competent ML practitioners.
For very large companies we often had fewer estimates due to data
collection limitations. Future research could incorporate broader
LinkedIn data sweeps and public outputs such as papers or code
repositories to assess competence further. 

\subsection{Conclusion}\label{h.luj6gsdcoues}

We identify a substantial pool of technically competent ML Research Talent
in the low thousands across companies offering ML consulting services.
These organizations represent a viable path for expanding capacity for
technical AI assurance work.

\vspace{12pt}
\noindent\rule{\linewidth}{0.4pt}
\vspace{6pt}

{\small\sffamily\bfseries Conflicts of Interest}\\
{\small The authors report no conflict of interest.}

\vspace{6pt}
{\small\sffamily\bfseries Author Contributions}\\
{\small Conceptualization: MS, NZ, RB, GL, SH $\cdot$
Methodology: MS, NZ, OE, GL $\cdot$
Software: FA-Z, OE $\cdot$
Formal analysis: MS, FA-Z, OE $\cdot$
Investigation: RB, NZ $\cdot$
Data curation: RB, MS $\cdot$
Visualization: MS, OE $\cdot$
Writing, original draft: MS $\cdot$
Project administration: MS, RB $\cdot$
Supervision: MS, SH}

\vspace{6pt}
{\small\sffamily\bfseries Acknowledgements}\\
{\small The authors acknowledge Coefficient Giving who funded this study and Lucas Sato, technical staff at METR, who helped run AI agent work trial comparison benchmarks.}

\vspace{6pt}
{\small\sffamily\bfseries Statement on AI Use}\\
{\small We used large language models (OpenAI GPT-4/5, Google Gemini 2.5 Flash/Pro, and Anthropic Claude 3.7/4/4.5) as described in the method section. Beyond individual prompts, models were used for literature search and to generate early section drafts. Models were also used for data-analysis support. All AI-generated content was thoroughly reviewed, verified, and edited by the authors, who take full responsibility for the final content.}

\vspace{6pt}
\noindent\rule{\linewidth}{0.4pt}

\clearpage
\begin{figure}[p]
\centering
\vspace*{\fill}
\includegraphics[width=\linewidth,keepaspectratio]{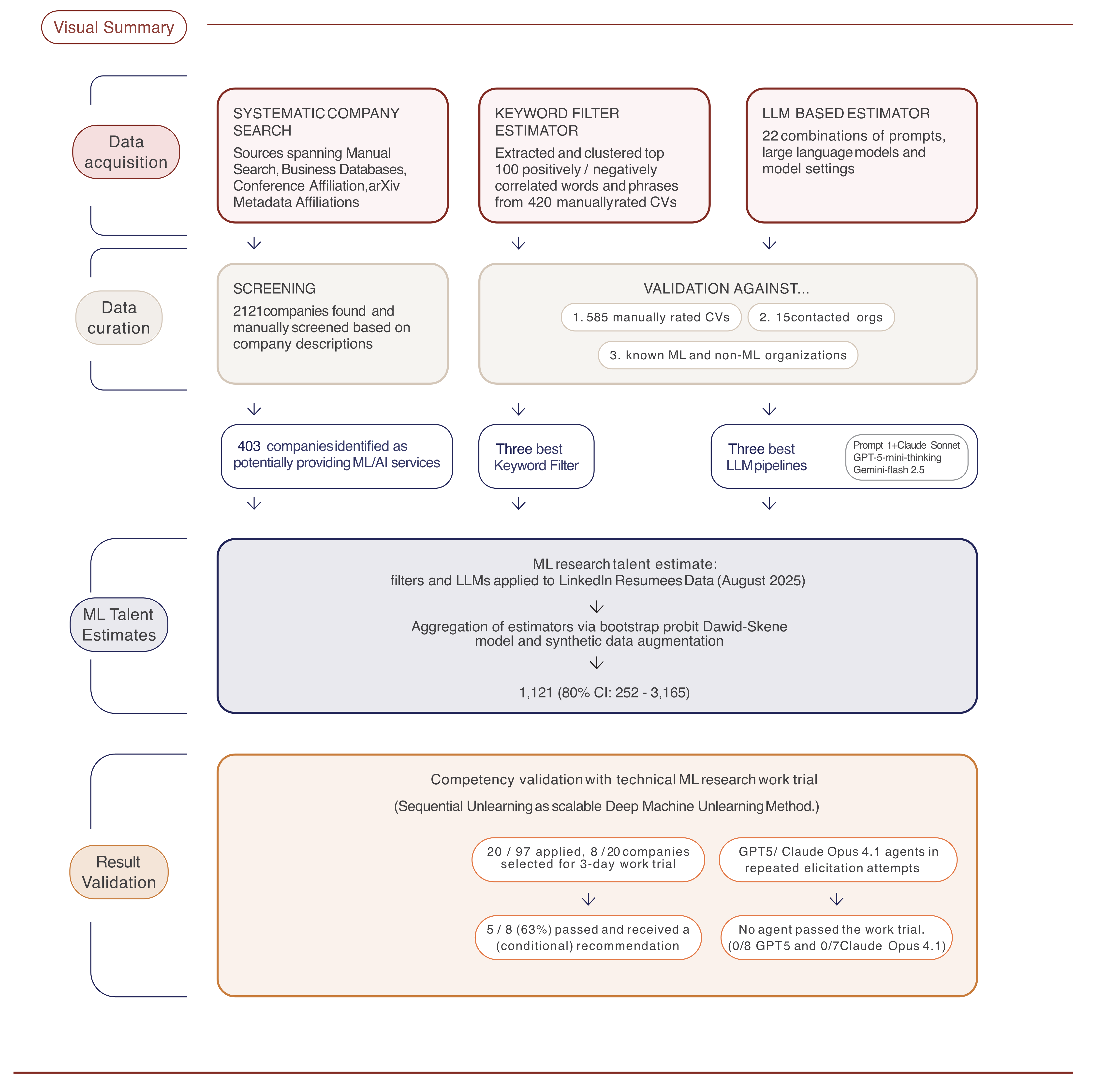}
\caption*{Visual summary of the study}
\label{fig:visual-summary}
\vspace*{\fill}
\end{figure}

\clearpage
\section{Introduction}\label{h.9jv9vou6zpxp}

The bottleneck for AI safety and alignment work is access to technical
talent. Capital expenditure for AI infrastructure is rising at an
exceptional rate (see [1--4]), yet outside a handful of frontier labs
and academic groups, few organizations can execute difficult alignment
and evaluation work at production speed. Existing programs such as
MATS (\href{https://www.matsprogram.org/}{ML
Alignment \& Theory Scholars}) have supported several hundred scholars
since 2021---meaningful, but small relative to the scale of the challenge.

This paper tests whether IT consulting firms can supply a scalable
pool of competent engineers and researchers for technical AI assurance.
The sector is sizeable, with individual companies employing hundreds of
thousands of workers and investing heavily in AI capabilities---for
example Accenture's \$3\,billion plan and Capgemini's multi‑year
multibillion‑euro initiative
(\href{https://newsroom.accenture.com/news/2023/accenture-to-invest-3-billion-in-ai-to-accelerate-clients-reinvention}{Accenture
Newsroom}, \href{https://www.capgemini.com/news/press-releases/capgemini-announces-augmented-engineering-offerings-powered-by-gen-ai/}{Capgemini}).
If a fraction of this workforce can be identified and directed toward
alignment tasks, funders and operators could expand capacity rapidly.

We contribute three things. First, we assemble a first-of-its-kind
systematic sample of ML consultancies globally; second, we estimate
their technical ML research headcount using LinkedIn resumes; and third,
we validate capabilities through a multi-day work trial benchmarked
against modern LLM agents and established AI/non-AI companies.

\clearpage
\section{Results}\label{h.n240xtg705e7}

\subsection{Systematic Search}\label{h.9e40uokxx3pv}

Figure 1 provides a flow chart of the inclusion and exclusion of
companies at various stages of the project. In total 2\,121 companies were
identified and screened, eventually leaving 403 companies that promised
to provide ML consultancy services. Consultancies were globally
distributed, but leaned towards the US and Europe as shown in Figure 2.

The total headcount of current employees based on LinkedIn Sales
Navigator was 3\,269\,000 with the top organizations having hundreds of
thousands of employees. In 126/403 (31.3\%) of companies we were not
able to detect any talent as per our definition.

\begin{figure}[htbp]
\centering
\pandocbounded{\includegraphics[keepaspectratio]{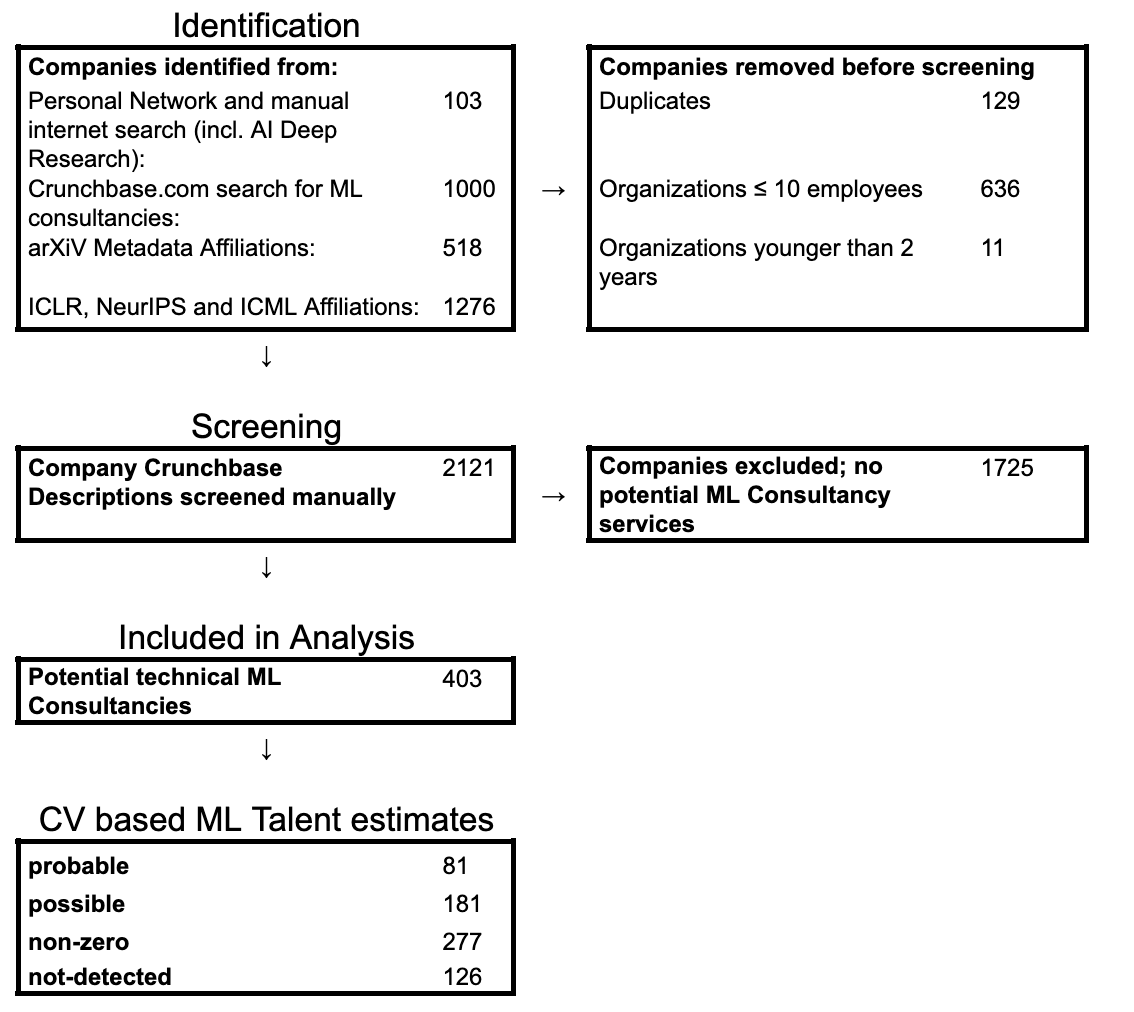}}
\caption{Flow chart of search process}
\label{fig:flowchart}
\end{figure}

\begin{figure}[htbp]
\centering
\pandocbounded{\includegraphics[keepaspectratio]{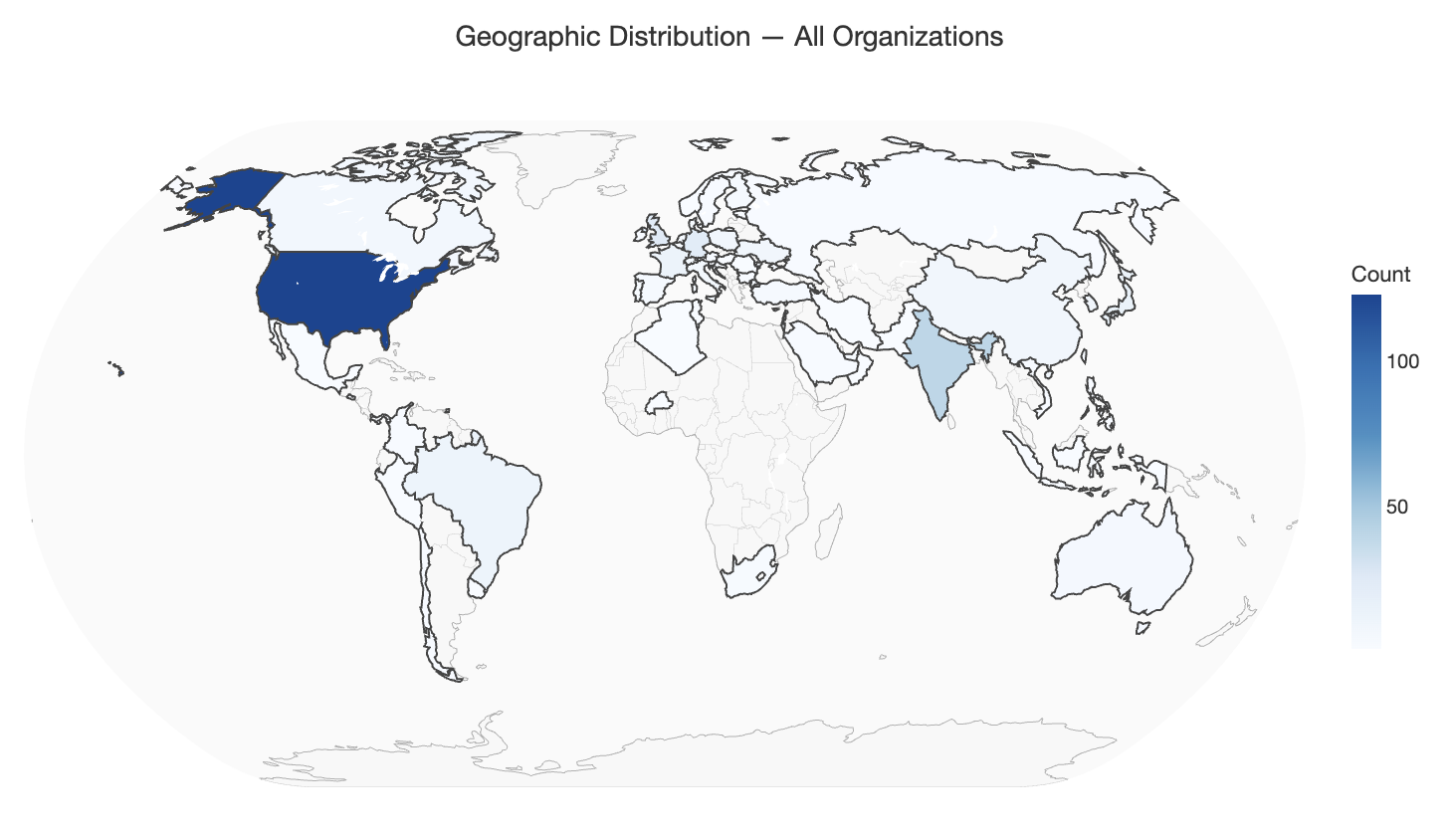}}
\pandocbounded{\includegraphics[keepaspectratio]{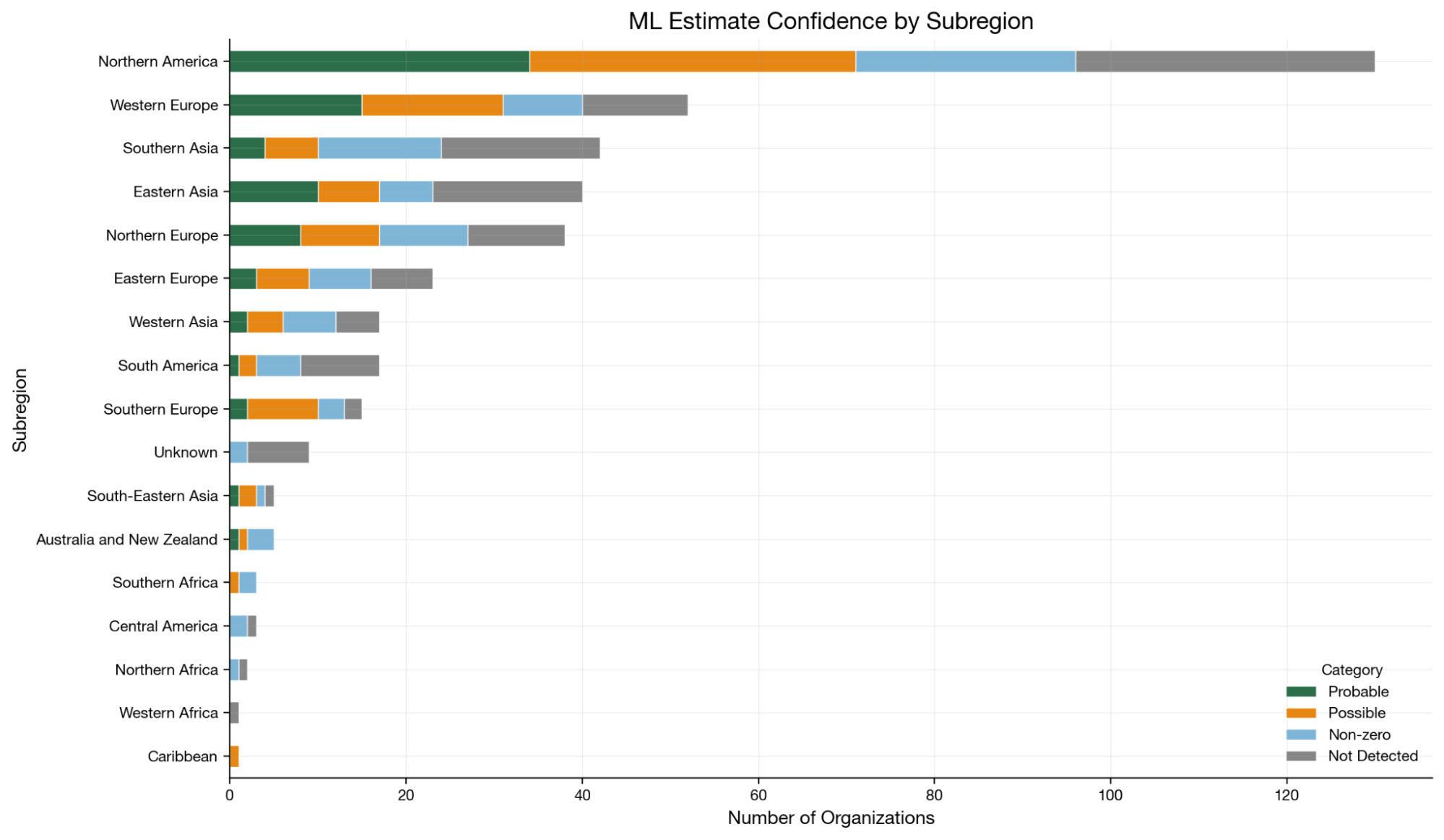}}
\caption{A) Map of 403 identified ML Consultancies B) Breakdown by Country and category. Organizations are classified into mutually exclusive confidence categories based on their ML research talent estimate distributions (q10, q50, q90 representing the 10th, 50th, and 90th percentiles).}
\label{fig:map}
\end{figure}

\clearpage
\scriptsize
\renewcommand{\arraystretch}{0.95}
\begin{longtable}[]{@{}
  >{\raggedright\arraybackslash}p{(\linewidth - 8\tabcolsep) * \real{0.32}}
  >{\raggedright\arraybackslash}p{(\linewidth - 8\tabcolsep) * \real{0.17}}
  >{\raggedright\arraybackslash}p{(\linewidth - 8\tabcolsep) * \real{0.17}}
  >{\raggedright\arraybackslash}p{(\linewidth - 8\tabcolsep) * \real{0.17}}
  >{\raggedright\arraybackslash}p{(\linewidth - 8\tabcolsep) * \real{0.17}}@{}}
\caption{Descriptive Statistics on various cohorts of the analysis}\label{tab:descriptive}\\\\
\toprule\noalign{}
\endhead
\bottomrule\noalign{}
\endlastfoot
\textbf{Characteristic} & \textbf{All} & \textbf{Probable} & \textbf{Possible} & \textbf{Non-zero} \\
{Total} & {} & {} & {} & {} \\
{~ ~ Organization N} & {403 (100.0\%)} & {81 (100.0\%)} & {100
(100.0\%)} & {96 (100.0\%)} \\
{~ ~ Total employees} & {3 269 000} & {2 931 516} & {280 284} & {3
365} \\
{~ ~ Median founding year} & {2014} & {2014} & {2013} & {2016} \\
{~ ~ Median total employees} & {28} & {442} & {90} & {22} \\
{~ ~ ML research staff (q50)} & {1 121 (252 - 3 165)} & {890 (218 - 2 522)} &
{205 (33 - 537)} & {24 (0 - 99)} \\
{~ ~ ML \% of total} & {0.0\% (0.0\% - 0.1\%)} & {0.0\% (0.0\% - 0.1\%)}
& {0.1\% (0.0\% - 0.2\%)} & {0.7\% (0.0\% - 2.9\%)} \\
\midrule
{Small (\textless{} 100 employees)} & {} & {} & {} & {} \\
{~ ~ Organization N} & {284 (70.5\%)} & {19 (23.5\%)} & {55 (55.0\%)} &
{87 (90.6\%)} \\
{~ ~ Median total employees} & {15} & {48} & {36} & {20} \\
{~ ~ ML research staff (q50)} & {166 (28 - 420)} & {71 (20 - 140)} & {71 (8 -
179)} & {23 (0 - 94)} \\
{~ ~ ML \% of total} & {2.6\% (0.4\% - 6.5\%)} & {6.9\% (1.9\% -
13.6\%)} & {2.9\% (0.4\% - 7.4\%)} & {1.0\% (0.0\% - 4.1\%)} \\
\midrule
{Medium (100-999 employees)} & {} & {} & {} & {} \\
{~ ~ Organization N} & {76 (18.9\%)} & {37 (45.7\%)} & {30 (30.0\%)} &
{9 (9.4\%)} \\
{~ ~ Median total employees} & {232} & {388} & {215} & {112} \\
{~ ~ ML research staff (q50)} & {398 (134 - 823)} & {320 (117 - 638)} & {76
(16 - 180)} & {1 (0 - 4)} \\
{~ ~ ML \% of total} & {1.7\% (0.6\% - 3.5\%)} & {2.1\% (0.8\% - 4.2\%)}
& {1.0\% (0.2\% - 2.5\%)} & {0.1\% (0.0\% - 0.4\%)} \\
\midrule
{Large (1,000-9,999 employees)} & {} & {} & {} & {} \\
{~ ~ Organization N} & {23 (5.7\%)} & {10 (12.3\%)} & {11 (11.0\%)} & {0
(0.0\%)} \\
{~ ~ Median total employees} & {2 900} & {4 600} & {2 300} & {} \\
{~ ~ ML research staff (q50)} & {154 (32 - 446)} & {117 (25 - 337)} & {37 (6
- 108)} & {0 (0 - 0)} \\
{~ ~ ML \% of total} & {0.2\% (0.0\% - 0.5\%)} & {0.2\% (0.1\% - 0.7\%)}
& {0.1\% (0.0\% - 0.4\%)} & {n/a} \\
\midrule
{Giant ($\geq$10\,000 employees)} & {} & {} & {} & {} \\
{~ ~ Organization N} & {20 (5.0\%)} & {15 (18.5\%)} & {4 (4.0\%)} & {0
(0.0\%)} \\
{~ ~ Median total employees} & {60 000} & {85 000} & {42 500} & {} \\
{~ ~ ML research staff (q50)} & {401 (57 - 1 474)} & {381 (55 - 1 406)} & {20
(2 - 68)} & {0 (0 - 0)} \\
{~ ~ ML \% of total} & {0.0\% (0.0\% - 0.0\%)} & {0.0\% (0.0\% - 0.0\%)}
& {0.0\% (0.0\% - 0.0\%)} & {n/a} \\
\midrule
{Regions (orgs)} & {} & {} & {} & {} \\
{~ ~ Northern America} & {130 (32.3\%)} & {34 (42.0\%)} & {37 (37.0\%)}
& {25 (26.0\%)} \\
{~ ~ Western Europe} & {52 (12.9\%)} & {15 (18.5\%)} & {16 (16.0\%)} &
{9 (9.4\%)} \\
{~ ~ Southern Asia} & {42 (10.4\%)} & {4 (4.9\%)} & {6 (6.0\%)} & {14
(14.6\%)} \\
{~ ~ Eastern Asia} & {40 (9.9\%)} & {10 (12.3\%)} & {7 (7.0\%)} & {6
(6.2\%)} \\
{~ ~ Northern Europe} & {38 (9.4\%)} & {8 (9.9\%)} & {9 (9.0\%)} & {10
(10.4\%)} \\
{~ ~ Eastern Europe} & {23 (5.7\%)} & {3 (3.7\%)} & {6 (6.0\%)} & {7
(7.3\%)} \\
{~ ~ Western Asia} & {17 (4.2\%)} & {2 (2.5\%)} & {4 (4.0\%)} & {6
(6.2\%)} \\
{~ ~ South America} & {17 (4.2\%)} & {1 (1.2\%)} & {2 (2.0\%)} & {5
(5.2\%)} \\
{~ ~ Southern Europe} & {15 (3.7\%)} & {2 (2.5\%)} & {8 (8.0\%)} & {3
(3.1\%)} \\
{~ ~ Unknown} & {9 (2.2\%)} & {0 (0.0\%)} & {0 (0.0\%)} & {2 (2.1\%)} \\
{~ ~ Australia and New Zealand} & {5 (1.2\%)} & {1 (1.2\%)} & {1
(1.0\%)} & {3 (3.1\%)} \\
{~ ~ South-Eastern Asia} & {5 (1.2\%)} & {1 (1.2\%)} & {2 (2.0\%)} & {1
(1.0\%)} \\
{~ ~ Central America} & {3 (0.7\%)} & {0 (0.0\%)} & {0 (0.0\%)} & {2
(2.1\%)} \\
{~ ~ Southern Africa} & {3 (0.7\%)} & {0 (0.0\%)} & {1 (1.0\%)} & {2
(2.1\%)} \\
{~ ~ Northern Africa} & {2 (0.5\%)} & {0 (0.0\%)} & {0 (0.0\%)} & {1
(1.0\%)} \\
{~ ~ Western Africa} & {1 (0.2\%)} & {0 (0.0\%)} & {0 (0.0\%)} & {0
(0.0\%)} \\
{~ ~ Caribbean} & {1 (0.2\%)} & {0 (0.0\%)} & {1 (1.0\%)} & {0
(0.0\%)} \\
\end{longtable}
\normalsize
\renewcommand{\arraystretch}{1.2}

\clearpage

\begin{figure}[htbp]
\centering
\pandocbounded{\includegraphics[keepaspectratio]{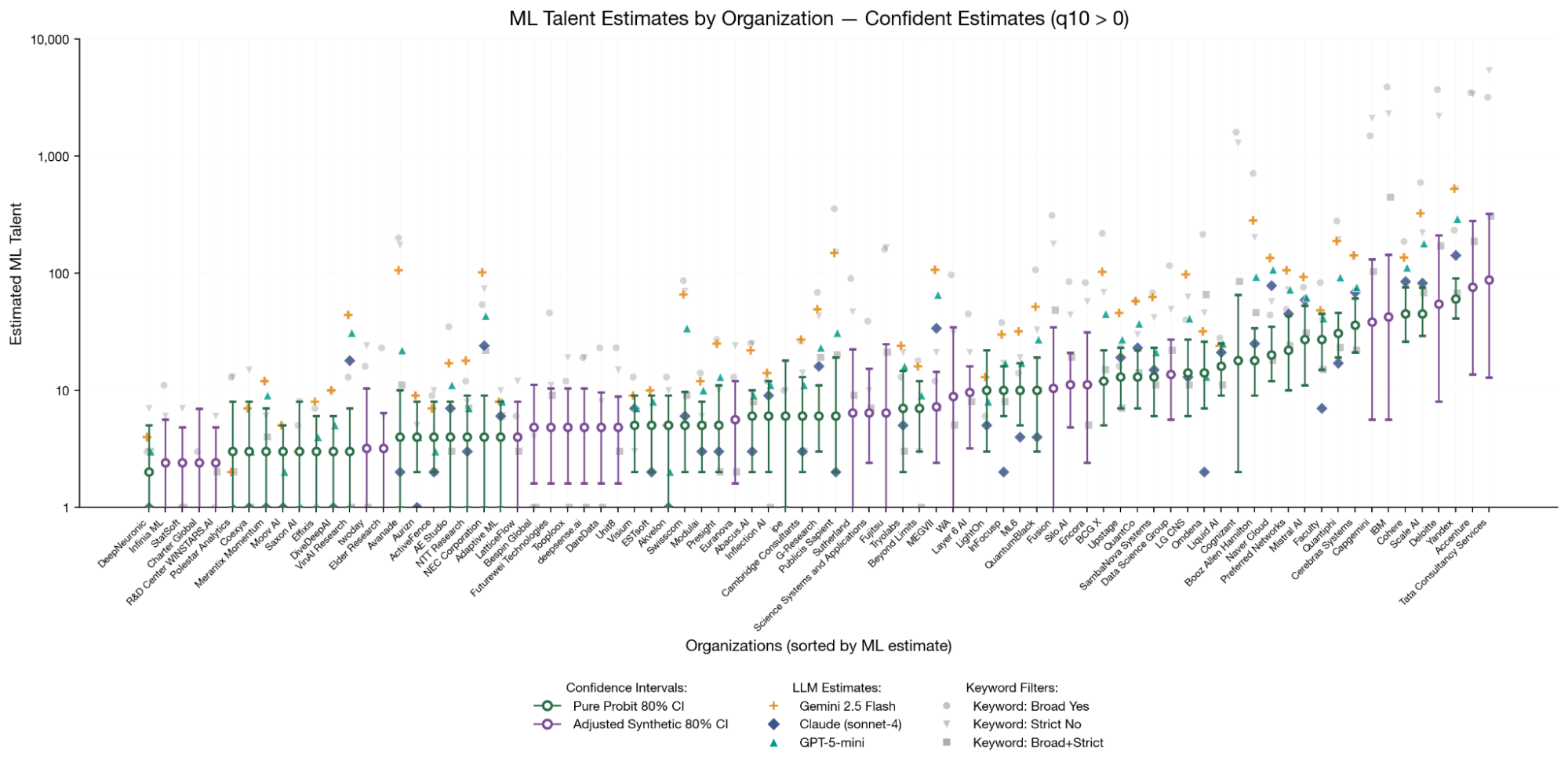}}
\caption{Individual and aggregated ML Research Talent estimates of 81 organizations among identified technical ML consultancies classified as probable---the 80\% confidence interval excludes zero, indicating confident ML presence. When LLM estimates were not available, synthetic data approaches were used (purple).}
\label{fig:ml-estimates}
\end{figure}

\begin{figure}[htbp]
\centering
\pandocbounded{\includegraphics[keepaspectratio]{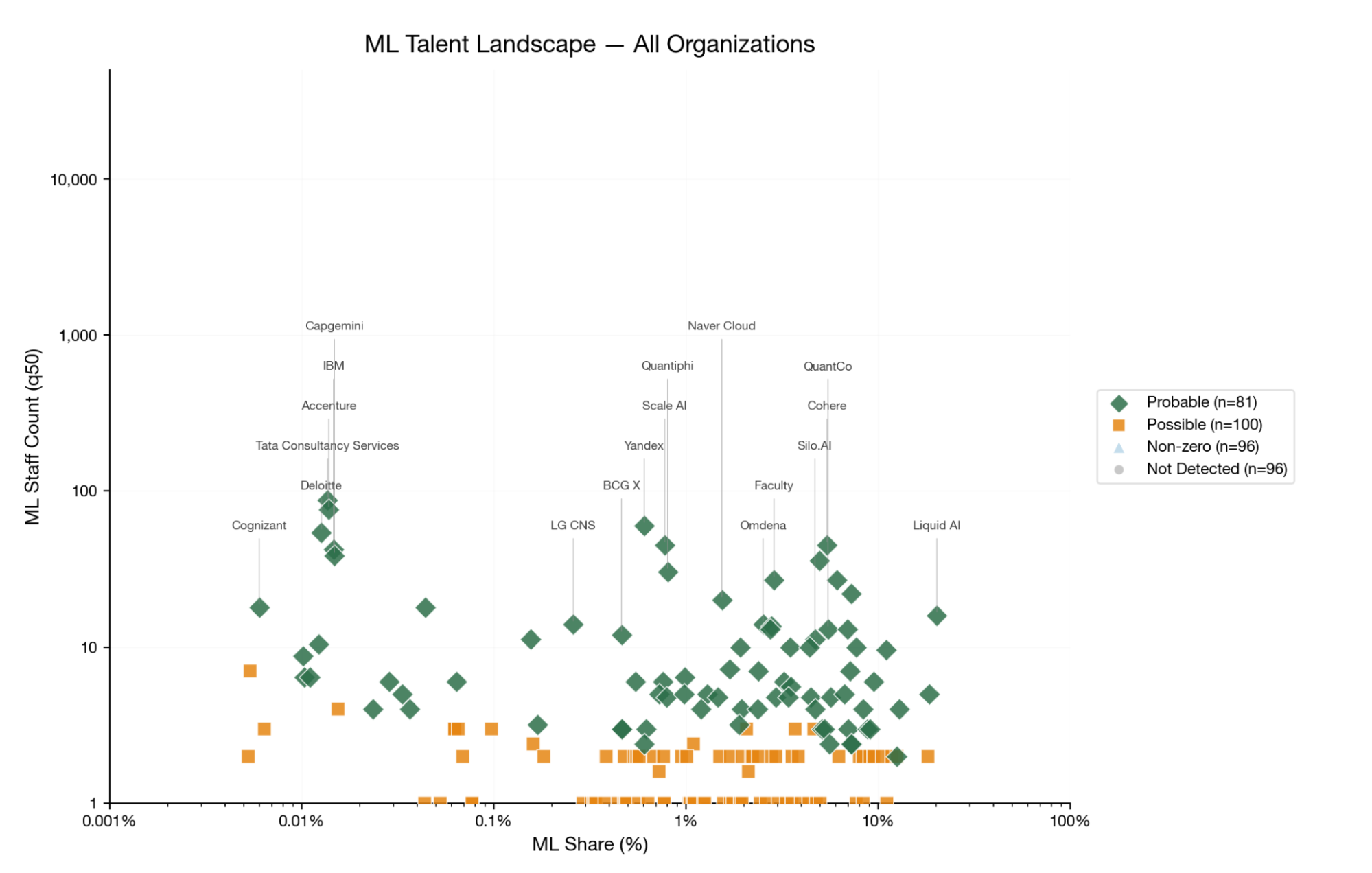}}
\caption{Visualization of ML staff count by share of ML Research Talent of total company employees. A) IT Consultancies B) 18 comparator ML Companies C) 18 non-ML companies}
\label{fig:ml-staff}
\end{figure}

\begin{figure}[htbp]
\centering
\pandocbounded{\includegraphics[keepaspectratio]{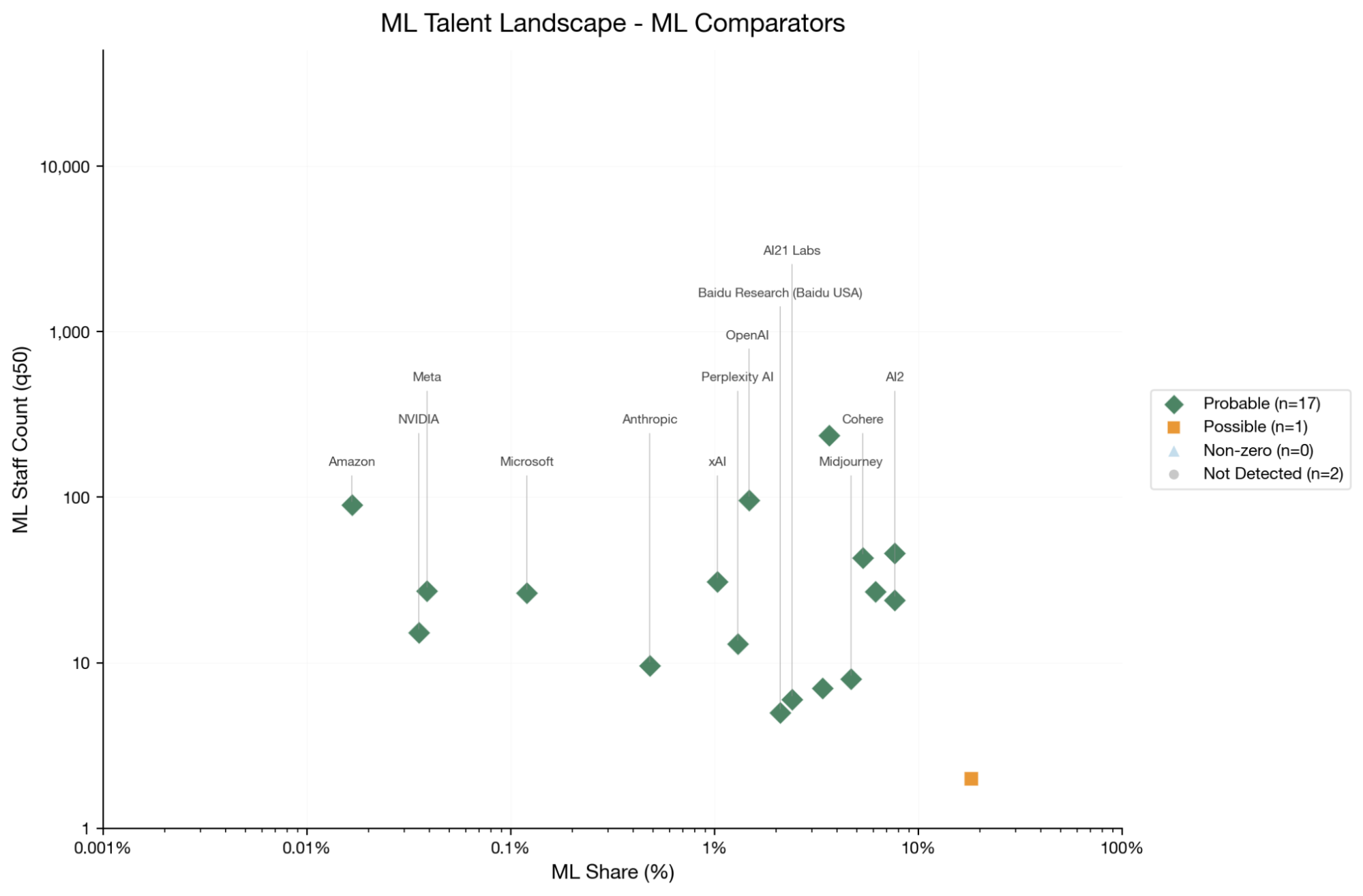}}
\pandocbounded{\includegraphics[keepaspectratio]{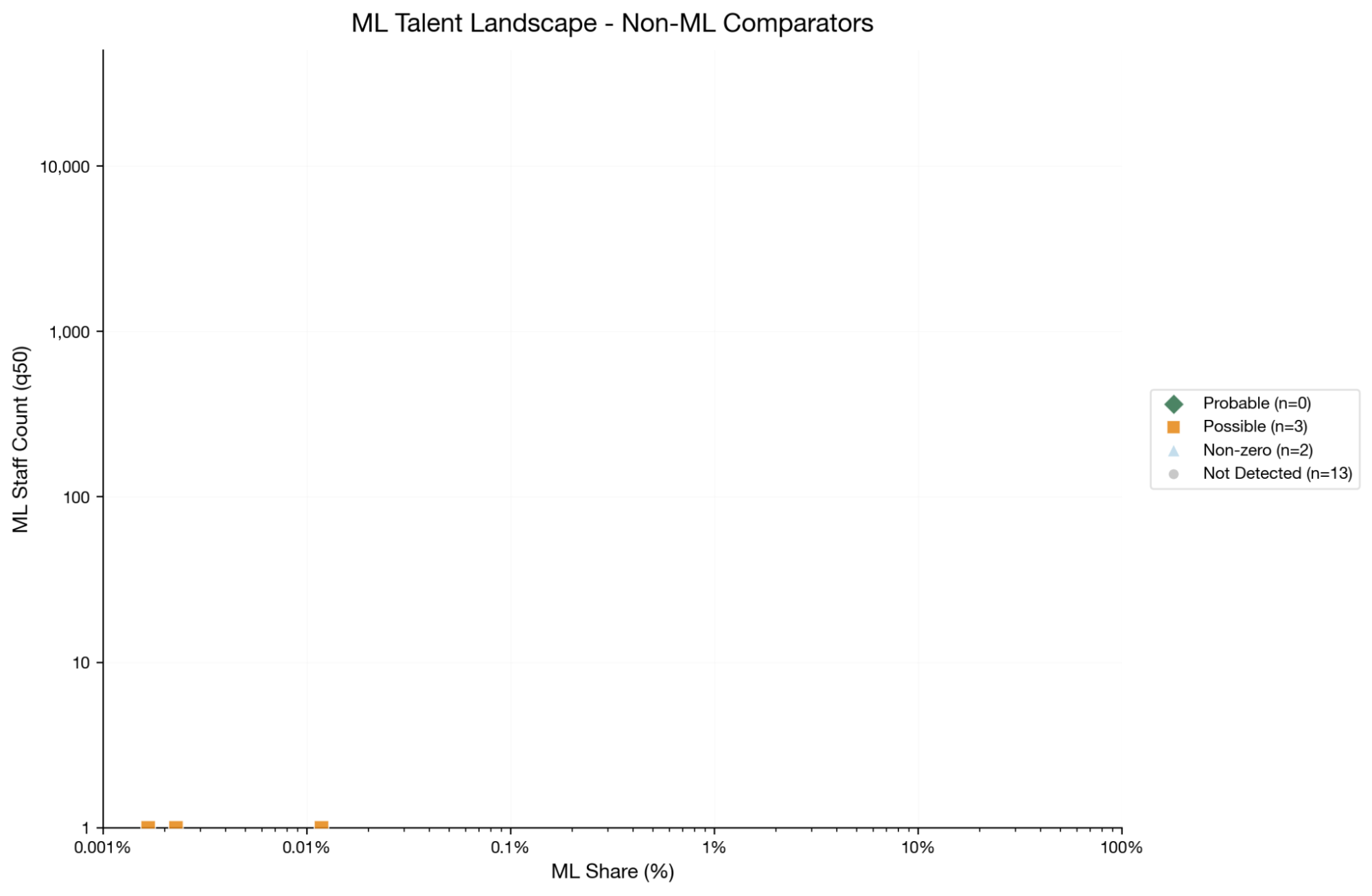}}
\caption{ML Research Talent estimates comparison}
\label{fig:estimates-comparison}
\end{figure}

\subsection{ML Research Talent Estimates}\label{h.4fw0wj8s1r4k}

Table~\ref{tab:descriptive} presents the aggregated ML Research Talent
estimates. For 81/403 (20\%) organizations the 80\% confidence interval
excludes zero, indicating relative confidence for ML Research Talent
presence; these firms account for 890 (218--2\,522), or 79.4\% of all
ML Research Talent in our sample (Figure 3). Figure 4 plots the ML
staff count against the percentage of technical ML Research Talent.
Individual company estimates and categorizations are available in the
Supplements.

Our final estimator had a sensitivity of 0.79 and specificity of 0.926,
yielding the best accuracy across estimators (0.89), a Positive
Likelihood Ratio (LR+) of 10.67, and a Negative Likelihood Ratio (LR-)
of 0.23.

As an additional validation step, we applied the selected estimation
methods to companies we knew to have high technical ML Research Talent counts and
to organizations who certainly don't. The results are presented in the
Supplements. Our method arrived at high technical ML Research Talent counts for
established AI organizations (e.g., OpenAI, Mistral AI, HuggingFace) and
low to zero counts for non-AI companies (e.g., Patagonia, Crocs, Inc.,
The British Museum) when priors were chosen accordingly.

However, for several established ML organizations---particularly those where LLM-based estimates were unavailable and synthetic imputation was used instead---the pipeline produced estimates that appear implausibly low. For example, Anthropic (2\,000 employees) received a median estimate of only 9 (1--26) ML research staff. Similarly, other large organizations relying on synthetic estimates (marked with * in the Supplements), such as Amazon, Meta, Microsoft, and NVIDIA, yielded lower estimates relative to their known ML research activity. This pattern was most pronounced for companies where the consultancy-calibrated priors (Table~\ref{tab:betapriors}) were applied to organizations with fundamentally different talent compositions, and where the absence of LLM-based individual CV assessments left the pipeline reliant solely on keyword-derived synthetic annotations.

\subsection{Work Trials}\label{h.z6symiif4y2g}

In parallel to the talent estimation work, we reached out to a total of
97/403 (24.1\%) companies for a 2-month engagement, including a 3-day work
trial. 57/97 (59.8\%) were not part of the IT consultancies we
identified as ``ML consultancies'' and functioned as additional
validation. We were able to get in contact with 47/97 (48.5\%) companies
and eventually received 20/97 (20.6\%) applications. 12 organizations
were either rejected or pulled back during the application process,
leaving 8 organizations who were offered a work trial. All organizations
invited to a work trial performed it during a 4 week period in July to
August 2025. 

The work trial task was to implement a "Sequential Unlearning" method -
a multi-stage wrapper around their existing RMU algorithm that
progressively unlearns data in folds - and integrate it into their
evaluation codebase, with daily progress updates and a final
code/write-up deliverable. The full repository is available in the
appendix.

The breakdown of the work trial evaluation is presented in Table~\ref{tab:worktrial}.
Three organizations received a recommendation, two a conditional
recommendation, 3/8 (37.5\%) no recommendation. Prices varied from
\$45 to \$350 per hour, with two organizations providing the work trial
free of charge. For confidentiality reasons, individual results,
evaluations and company names related to the work trial are not shared
with this publication.

In collaboration with a member of the technical staff from METR we
let GPT-5 and Claude Opus 4.1 Agents perform multiple runs at the
work trial task as well. After some elicitation, agents were evaluated
by the same criteria. All agents scored between 30 and 40\% on our
work trial, resulting in no recommendation.

\scriptsize
\begin{longtable}[]{@{}
  >{\raggedright\arraybackslash}p{(\linewidth - 6\tabcolsep) * \real{0.2500}}
  >{\raggedright\arraybackslash}p{(\linewidth - 6\tabcolsep) * \real{0.2500}}
  >{\raggedright\arraybackslash}p{(\linewidth - 6\tabcolsep) * \real{0.2500}}
  >{\raggedright\arraybackslash}p{(\linewidth - 6\tabcolsep) * \real{0.2500}}@{}}
\caption{Overview of work trial results. Most agents implemented a somewhat well-documented Sequential Unlearning, integrated it into the pipeline, and provided usable configs with coherent write-ups. However, only a small subset executed any unlearning to completion, and none followed through with RTT evaluation.}\label{tab:worktrial}\\\\
\toprule\noalign{}
\endhead
\bottomrule\noalign{}
\endlastfoot
{} & \textbf{No Recommendation (\textless50\%)} & \textbf{Conditional Recommendation (50-70\%)} & \textbf{Recommendation (\textgreater70\%)} \\
{ML Consultancies} & {3/8 (37.5\%)} & {2/8 (25\%)} & {3/8 (37.5\%)}

{Note: all three organizations scored \textgreater90\%} \\
{AI Agents} & \begin{minipage}[t]{\linewidth}\raggedright
{ChatGPT-5: 8/8 (100\%)\\
Claude Opus 4.1: 8/8 (100\%)}\strut
\end{minipage} & {-} & {-} \\
\end{longtable}
\normalsize
\renewcommand{\arraystretch}{1.2}

\clearpage
\section{Discussion}\label{h.muux9icv0fru}

To our knowledge, this is the first systematic assessment of the
``dormant'' pool of technical ML Research Talent inside IT
consultancies. We introduce a transparent estimation pipeline for
identifying high‑quality ML practitioners, and we validate it via
multiple modalities: validation datasets for screening and estimation,
organization‑level outreach, and targeted work trials. While our work
was motivated by advancing AI assurance work, these methods offer a
pragmatic general map of where skilled ML capacity might exist.

Our global sweep estimated roughly 1\,100 individuals with robust AI/ML
profiles across 403 consultancies. Final ML Research Talent
estimates aligned with our work-trial experiences and conversations with
various IT consultants. Organizations concentrated in the US and Europe.
Diversified conglomerates whose services span far beyond ML/AI are one
of the reasons why overall ML Research Talent density with \textasciitilde0.01\%
is low. At the level of individual firms, however, the picture is
heterogeneous, with skilled technical ML staff ranging up to 20\%.
Funders and program managers should therefore avoid treating
``consultancies'' as a homogeneous class and require careful targeting
of the right sub‑units.

AI agents are continuously advancing, but at least in our analysis they
fell far behind consultancies, despite repeated expert elicitation.
This suggests that the right consultancies can deliver useful ML R\&D
outputs we can't replace with AI today. Activation requires deliberate
scoping, credible sponsorship, and (often) smaller, faster contracting
paths.

Our findings should be interpreted with several caveats. First, we
used a particular definition of ML Research Talent that could exclude
competent practitioners in adjacent domains such as ML Ops. The systematic
search was English‑only and anchored to Crunchbase and LinkedIn, likely
underrepresenting non‑Western markets and firms with limited public
footprints. Our final estimators, with a positive likelihood ratio of
\textasciitilde11 and a negative likelihood ratio of 0.23, remain an
imperfect signal, particularly at the scale of over 3 million assessed
individuals.

For several established ML organizations---where LLM-based estimates were
unavailable and synthetic imputation was used instead---the pipeline
produced implausibly low estimates (e.g., Anthropic: 9 of 2\,000
employees). This reflects a mismatch between our consultancy-calibrated
priors (Table~\ref{tab:betapriors}) and the fundamentally higher talent
density at frontier labs. Future applications to such organizations would
benefit from domain-specific priors or full LLM-based CV evaluation.

Our work trial was an important
validation step, but resumes remain an inherently noisy proxy for
competence and we were not able to systematically connect individual CVs
to public artifacts (e.g., publications, GitHub/GitLab) at scale.
The work‑trial component involved only 20 organizations, a single
3‑day task, and no true control arm of lab/academic researchers. We
instead leveraged state‑of‑the‑art AI coding tools. Through conducting
the work trials a consistent operational lesson also emerged: even
highly competent consultancies prefer an external ``vision holder''
(e.g., a senior AI‑alignment researcher) to specify outcomes and own the
research direction. Smaller organizations we engaged tended to cite
capacity challenges, larger organizations minimum contract sizes.

Within these limits, we find clear ``proof of existence'' for technical
ML practitioners who can execute a challenging technical ML task to the
highest levels of satisfaction. This should be an encouragement for
funders and program managers in the AI assurance and general AI space to
consider this accelerated path of increasing capacity for AI safety
projects.

\clearpage
\section{Methods}\label{h.wujmwe8ddwwx}

\subsection{Systematic company search}\label{h.5kqxyalyg3ob}

Over the course of June to August 2025 we identified companies via 

\begin{enumerate}
\tightlist
\item
  {personal network recommendations,}
\item
  {unstructured web search with 2 independent research assistants
  (leveraging Deep Research functionalities of LLMs),}
\item
  {extracting arXiv affiliation metadata,}
\item
  {extracting ICLR / ICML / NeurIPS affiliations back to 2019, and}
\item
  {Crunchbase.com database keyword search.}
\end{enumerate}

The details for the search strategy are available in the supplements
(see \hyperref[h.tjifkul59853]{Search Strategies}). All company
names identified in steps 1 to 4 were searched for and exported from
Crunchbase. We screened all company descriptions provided by Crunchbase
for relevance to technical ML consultancy services, using Claude Sonnet
3.7 and a human validator. Organizations with fewer than 10 staff members
on Crunchbase (and later LinkedIn exports), or younger than 2 years were
excluded. The date of Crunchbase data extraction for final analysis was
8th of August 2025.

There is an emerging literature on the use of LinkedIn for labor
measurement. LinkedIn has over 1.2 billion profiles and we expect it to
be the most representative data source available
(\href{https://www.linkedin.com/pulse/50-linkedin-statistics-every-professional-should-ti9ue/}{LinkedIn Statistics for Professionals} [8]).
Comparative studies show that LinkedIn‑derived indicators can correlate
well with official statistics while exhibiting selection effects by
country, age, and sector
(\href{https://epjdatascience.springeropen.com/articles/10.1140/epjds/s13688-021-00294-7}{SpringerOpen}, \href{https://academic.oup.com/jssam/article/12/5/1200/7728213}{Oxford
Academic}, \href{https://documents1.worldbank.org/curated/en/827991542143093021/pdf/World-Bank-Group-LinkedIn-Data-Insights-Jobs-Skills-and-Migration-Trends-Methodology-and-Validation-Results.pdf}{World
Bank}).

\subsection{Access to staff LinkedIn Resumes for evaluation}\label{h.7bdmlbeohfcj}

To assess a company's staff technical ML capabilities we decided to
evaluate respective LinkedIn resumes. We used two approaches: access via
LinkedIn's own Sales Navigator Recruiter Lite and Brightdata's LinkedIn
data. The former provided up-to-date information from LinkedIn, but did
only allow for simple filtering and keyword search. The latter enabled
us to screen CVs more comprehensively with LLMs, but did not always
represent current LinkedIn data and had to exclude extremely large
companies with more than 50,000 employees. 

\subsection{Methodology of Resume Evaluation Validation}\label{h.t2acsm4d9i87}

According to our data constraints we designed and validated two
complementary evaluation pipelines. One keyword based, another LLM
prompt based.

First, we created a comprehensive validation dataset with a total of
585 CVs extracted from LinkedIn. Two independent human
reviewers rated
all CVs against our definition of technical ML Research Talent as either yes or
no (see Supplements). The examples included randomly selected
individuals from IT consulting firms and work trial consultancies.
We also included random sets selected from small-scale technical AI
Alignment organizations and AI labs like OpenAI, Anthropic and Google to
ensure sufficient positive examples.

Based on 421 of the rated CVs we extracted the top 100 positively and
negatively associated keywords (see supplements). Features were selected
after a manual sanity pass to drop misleading tokens (e.g., organization
names or cities) at a ``strict'' and ``broad'' level for both positive
and negative terms. ``Broad'' selected terms with a
discriminative\_score \textgreater= 0.8 AND raw category specificity
\textgreater= 0.7, ``strict'' selected at a discriminative\_score
\textgreater= 0.9 AND raw category specificity \textgreater= 0.8. The
minimum frequency across all CVs for a keyword was five. In parallel we
created two evaluation LLM prompts
(see code repository)
we tested with different models and temperature settings at 0 or
default (usually 1.0 or 0.7; at the time of testing gpt-5 models did
not support a temperature parameter)\footnote{Temperature controls the randomness of an LLM's text generation. Lower values (e.g., 0.1) make outputs more deterministic and focused, while higher values (e.g., 0.9) make them more creative and unpredictable.}. Models
included:

\begin{itemize}
\tightlist
\item
  {Google: gemini-2.5-pro, gemini-2.5-flash and gemini-2.5-flash-lite}
\item
  {OpenAI: gpt-5-thinking-2025-08-07, gpt-5-mini-thinking-2025-08-07,
  gpt-5-nano-thinking-2025-08-07}
\item
  {Anthropic: opus-4-1-20250805, sonnet-4-20250514}
\end{itemize}

All permutations of keyword filters and all permutations of LLMs were
validated against the full 585 CVs. The top three performing filters and
top performing model per LLM provider were selected: broad\_yes,
strict\_no, and broad\_yes\_strict\_no for keywords. For LLMs, we
selected prompt 1 and gemini-2.5-flash, sonnet-4-20250514, and
gpt-5-mini-thinking-2025-08-07 as they provided robust results while
being relatively fast and cheap (see Table~\ref{tab:keywords} for keywords and prompt).
Exploratory results for worktrial companies aligned with both company
statements about their internal capacity as well as us reviewing CVs
manually.

\clearpage
\scriptsize
\begin{longtable}[]{@{}
  >{\raggedright\arraybackslash}p{(\linewidth - 4\tabcolsep) * \real{0.3333}}
  >{\raggedright\arraybackslash}p{(\linewidth - 4\tabcolsep) * \real{0.3333}}
  >{\raggedright\arraybackslash}p{(\linewidth - 4\tabcolsep) * \real{0.3333}}@{}}
\caption{Selected Keywords and prompts}\label{tab:keywords}\\\\
\toprule\noalign{}
\endhead
\bottomrule\noalign{}
\endlastfoot
\textbf{Highly Skilled technical ML Research Talent (Definition)} & \textbf{Keyword Filter} & \textbf{LLM Prompt} \\
\begin{minipage}[t]{\linewidth}\raggedright
{Professionals who can:}

\begin{itemize}
\tightlist
\item
  {Train models from scratch: Comfortable implementing and training
  transformer architectures like GPT-2 end-to-end, including loss
  calculation, attention mechanisms, and training loops}
\item
  {Work from specs to code: Take a method specification and build a
  working implementation, handling data pipelines, model internals, and
  distributed training setups}
\item
  {Debug model behavior: Diagnose and isolate root causes when training
  goes wrong or models behave unexpectedly}
\item
  {Engage with research: Read papers, understand novel approaches, and
  translate ideas into concrete implementation plans}
\item
  {Communicate clearly: Report progress, challenges, and solutions in
  accessible terms}
\item
  {Evidence: Public GitHub repos, papers/blog posts, or other forms of
  demonstrated experience }
\end{itemize}
\end{minipage} & \begin{minipage}[t]{\linewidth}\raggedright
{ML Selection}{\hfill\break
("machine learning" OR "machine‐learning" OR "ML" OR "deep learning" OR
"deep-learning", "reinforcement learning" OR "reinforcement-learning" OR
"RL") AND }

Broad\_yes (optional)

("augmented generation" OR "agent reinforcement" OR "mats scholar" OR
"mats" OR "research scientist" OR "evals" OR "interpretability" OR
"feature engineering" OR "research intern" OR "candidate" OR "graduate
research assistant" OR "science institute" OR "staff research scientist"
OR "doctor") 

Strict\_no (optional)

{NOT ("certificate" OR "programmer" OR "council" OR "companies" OR
"capital" OR "proven track record" OR "pilot" OR "money" OR "specialist"
OR "chief" OR "udemy" OR "track record" OR "customer" OR "management" OR
"today" OR "cross functional" OR "administrator" OR "excellence" OR
"commerce" OR "linkedin" OR "leader" OR "incident" OR "tier" OR "brand"
OR "investment" OR "hr" OR "sites" OR "offerings" OR "prior" OR
"centers" OR "advising" OR "certified information" OR "key
responsibilities" OR "master data" OR "anti" OR "deadlines" OR
"physiology" OR "carbon" OR "impacts" OR "certified machine" OR
"qualification")}\strut
\end{minipage} & {Question: Could this person design and implement
complex ML architectures from scratch (e.g., transformers, VAEs,
diffusion models) and is qualified for technical AI engineering or
research? Quick Check:}

GOOD FIT - Look for:

- Built neural networks or created novel architectures

- Deep understanding demonstrated through: custom loss functions,
attention mechanisms, or architecture modifications

- Advanced degree in ML WITH thesis/research on model architecture (not
just applications)

- Implemented training algorithms or frameworks from scratch (not using
existing libraries)

- LLM/RLHF experience - distinguish between using vs. building these
systems

NOT A FIT - Reject if they're primarily:

- Using pre-built models (even advanced ones like fine-tuning LLMs)

- Data Scientists (dashboards, analytics, A/B tests)

- MLOps/DevOps without architecture design

- Hardware optimization for ML (GPU programming) without model design

- Applied ML without evidence of understanding underlying
math/algorithms

{Output: [ACCEPT/REJECT]} \\
\end{longtable}
\normalsize
\renewcommand{\arraystretch}{1.2}

Data was collected in the time from 1st -- 15th of August. For the
keyword analysis, we manually selected the company name on LinkedIn's
Sales Navigator Recruiter Lite and pasted the respective keyword
combinations together with the shared search term ("machine learning" OR
"machine‐learning" OR "ML" OR "deep learning" OR "deep-learning",
"reinforcement learning" OR "reinforcement-learning" OR "RL"). The
number of total company and filter specific hits were manually
transferred into our database. LinkedIn's public interface rounds counts
(e.g., \textgreater1,000 results rounded; \textgreater100k rounded to
10k) and offers limited public API access, constraining programmatic
precision at large scales.

For the LLM evaluation, we sourced all company profiles via Bright Data
(\href{http://brightdata.com}{brightdata.com})
for organizations with fewer than 25,000 LinkedIn profiles in the
previous filter search. Data was processed by selected models with
Prompt 1 and results cleaned from hallucinated profiles. We ran the
evaluation of approximately 250\,000 profiles using the batching APIs of
the three models. LLM estimates were excluded if the headcount was off
by a factor of more than 3 compared to the LinkedIn Sales Navigator
Recruiter Lite data.

\subsection{Aggregation of individual estimates}\label{h.gtel51e8av36}

We estimated technical ML headcount using a bootstrap multivariate
probit model that combines six imperfect annotators: three keyword
filters (broad, strict, and combined) and three LLM classifiers (Gemini
2.5 Flash, GPT-5 Mini, Claude Sonnet 4) which each produce a binary
label for each employee (ML expert / not an ML expert). Our approach is
based on the Dawid-Skene model, but rather than assuming independent
annotator errors, our approach models correlated mistakes through a
multivariate normal latent structure, capturing, for instance, when
multiple LLMs fail on the same edge cases. The model assumes each
annotator's binary decision arises from thresholding a
latent continuous score in probit space (inverse standard normal CDF),
allowing us to model correlations in the underlying decision-making
process rather than just in the binary outcomes. The full pipeline is
illustrated in Figure 5 and code available in the repository.

From 585 manually-labeled validation CVs (153 ML experts, 432 non-ML),
we estimated each annotator's sensitivity and
specificity via confusion matrices, and their pairwise error
correlations via tetrachoric correlation, which estimates the
correlation between latent continuous variables underlying binary
annotations, consistent with the probit model's
assumption that binary labels arise from thresholding latent Gaussian
scores. (see Figure 5). The combined probit model achieved the highest
accuracy (0.89) with sensitivity of 0.79 and specificity of 0.93,
outperforming any individual annotator.

For inference, we compute the posterior probability that each employee
is a true ML expert given their pattern of annotations across all six
methods. The likelihood of observing a particular annotation pattern
under each true label class (ML expert vs. non-expert) is computed as a
multivariate normal orthant probability---the probability mass in the
region of latent space corresponding to that binary pattern. The
posterior probability is then obtained via Bayes'
theorem, combining these likelihoods with company-size-stratified Beta
priors on ML Research Talent prevalence.

\clearpage
\scriptsize
\renewcommand{\arraystretch}{1.4}
\begin{longtable}[]{@{}
  >{\raggedright\arraybackslash}p{(\linewidth - 6\tabcolsep) * \real{0.2500}}
  >{\raggedright\arraybackslash}p{(\linewidth - 6\tabcolsep) * \real{0.2500}}
  >{\raggedright\arraybackslash}p{(\linewidth - 6\tabcolsep) * \real{0.2500}}
  >{\raggedright\arraybackslash}p{(\linewidth - 6\tabcolsep) * \real{0.2500}}@{}}
\caption{Company-size-stratified Beta priors on ML Research Talent prevalence}\label{tab:betapriors}\\\\
\toprule\noalign{}
\endhead
\bottomrule\noalign{}
\endlastfoot
\textbf{Organization Type} & \textbf{Company Size} & \textbf{Beta Prior} & \textbf{Mean} \\
{Consulting \& Comparator ML Organisations} & {\textless{} 100} &
{Beta(2.4, 21.7)} & {10\%} \\
{Consulting \& Comparator ML Organisations} & {100--1,000} & {Beta(3.0,
57.0)} & {5\%} \\
{Consulting \& Comparator ML Organisations} & {1,000--10,000} &
{Beta(1.6, 154.8)} & {1\%} \\
{Consulting \& Comparator ML Organisations} & {\textgreater{} 10,000} &
{Beta(1.0,
999.0)}
& {0.1\%} \\
{Comparator Non-ML} & {All sizes} & {Beta(1.0, 9999.0)} & {0.01\%} \\
{All (unknown size)} & {Unknown} & {Beta(1.6, 154.8)} & {1\%} \\
\end{longtable}
\normalsize
\renewcommand{\arraystretch}{1.2}

For organizations with only aggregate LinkedIn keyword counts (due to
lack of access or substantial headcount discrepancies), we generate
synthetic employee-level annotations using a Gaussian copula that
preserves the estimated correlation structure between keyword filters
while matching company-level prevalences (aggregate counts divided by
total headcount). This approach assumes that the latent correlation
structure estimated from validation data generalizes across companies,
allowing us to create realistic synthetic annotation patterns that can
be processed by the same probit model as real employee-level data.
These synthetic annotations are used only when real employee-level
data is unavailable, enabling unified estimation across all
companies.
Synthetic estimates were adjusted by a factor of
0.5 based
on systematic comparison with real-data estimates for companies where
both were available (see Figure 6).

The bootstrap procedure (1,000 iterations) captures five sources of
uncertainty: (1) confusion matrix estimation from resampled validation
data, (2) prior uncertainty via Beta sampling, (3) within-company
sampling variation, (4) correlation structure uncertainty from
resampled test data, and (5) realization uncertainty through Bernoulli
draws of true labels. The bootstrap procedure works by running the full
pipeline many times, and for each source of uncertainty, resampling
(with replacement) the relevant population or re-drawing from the
relevant distribution. The resulting distribution yields point estimates
(means) and intervals (10th/90th percentiles) for company-level
headcounts.

We defined subsets for technical ML Research Talent-dense organizations as
follows: Organizations are classified into mutually exclusive confidence
categories based on their ML research talent estimate distributions (q10, q50,
q90 representing the 10th, 50th, and 90th percentiles):

\begin{itemize}[nosep]
\item \textbf{Probable:} q10 $>$ 0 --- The 80\% confidence interval excludes zero, indicating confident ML presence
\item \textbf{Possible:} q50 $>$ 0 and q10 = 0 --- The central estimate is positive but the confidence interval includes zero
\item \textbf{Non-zero:} q90 $>$ 0 and q50 = 0 --- Only the upper bound is positive; the central estimate is zero
\item \textbf{Not Detected:} q90 = q50 = q10 = 0 --- All estimates are zero, indicating no ML signal detected
\end{itemize}

Pure probit estimates are used when available; otherwise, adjusted
synthetic estimates are used.

\begin{figure}[htbp]
\centering
\pandocbounded{\includegraphics[keepaspectratio]{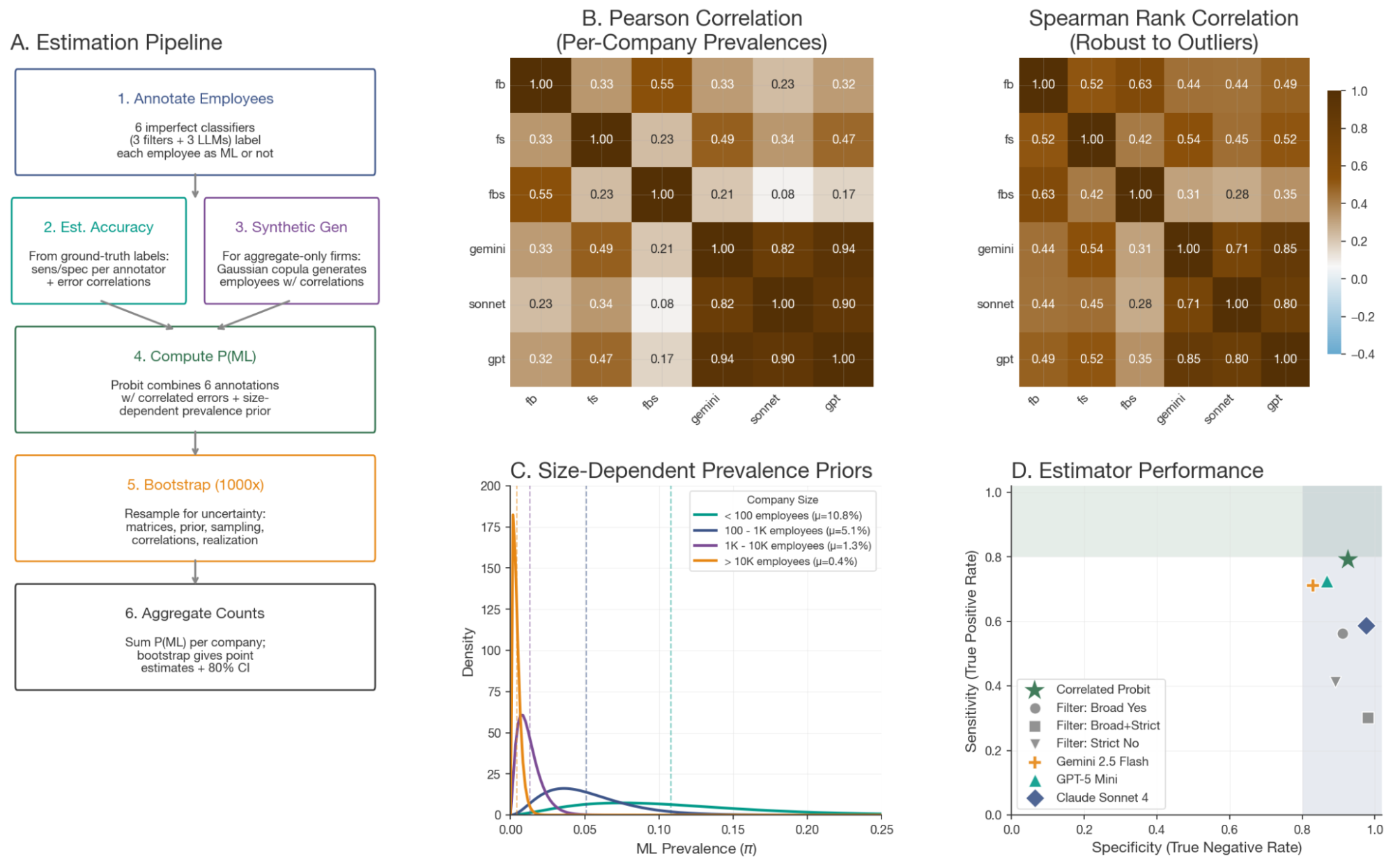}}
\caption{Comparison of Real and Synthetic estimates per company.}
\label{fig:real-synthetic}
\end{figure}

\begin{figure}[htbp]
\centering
\pandocbounded{\includegraphics[keepaspectratio]{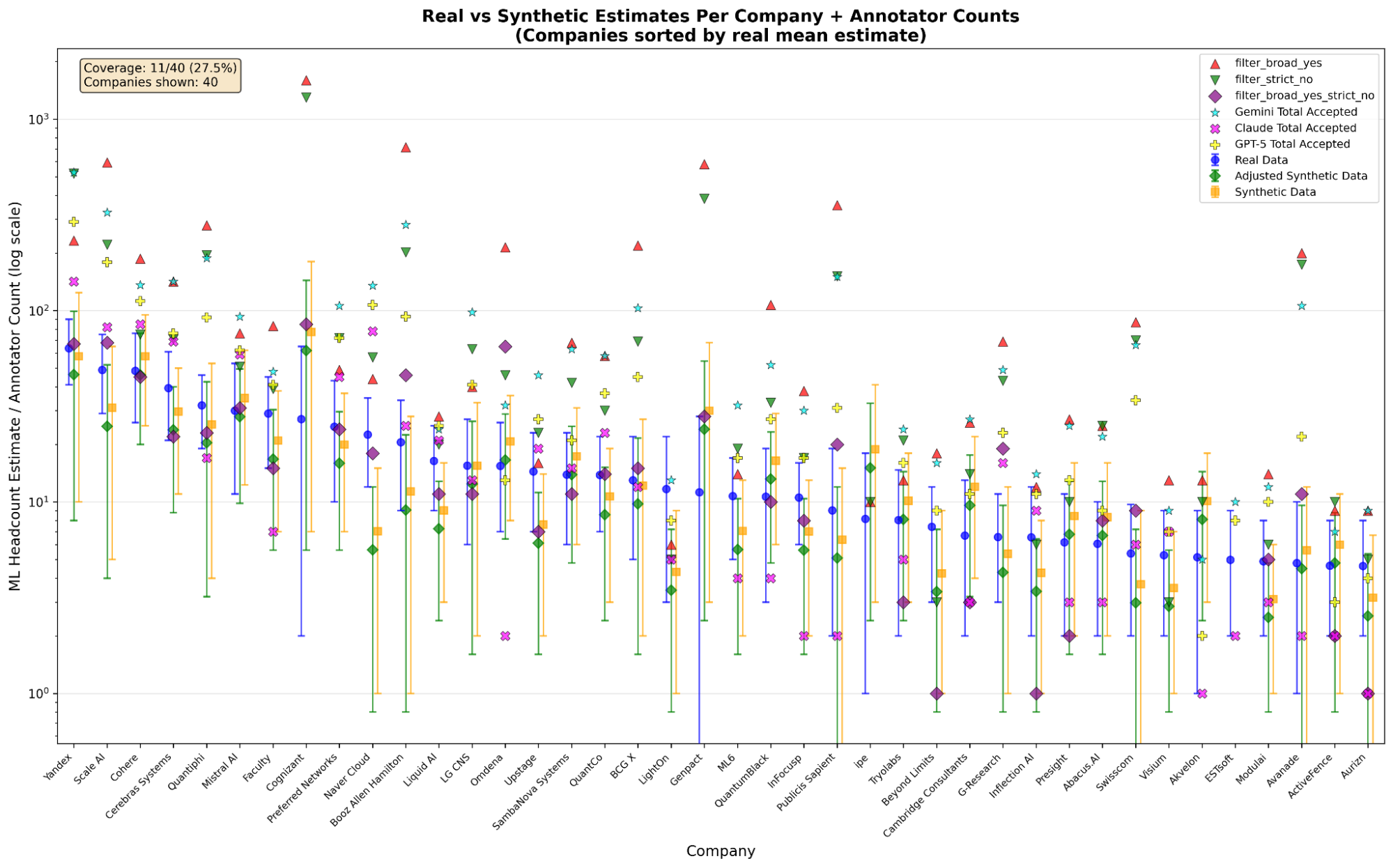}}
\caption{Methodology overview diagram. A) Pipeline overview B) Visualization of the distribution prior of the bootstrap probit Dawid-Skene model C) Sensitivity vs Specificity across all estimators, including final correlated probit.}
\label{fig:methodology}
\end{figure}

\clearpage
\subsection{Interviews and Work Trials}\label{h.bxilpniv77y5}

In parallel to the systematic search and ML Research Talent estimates, we
reached out to organizations for a work trial and consecutive two-month
technical AI alignment project to be funded by Coefficient Giving.

Companies were approached with a comprehensive project description.
Applications were scored by at least two reviewers for competence, scale
and maturity (see supplements). We followed up at least two times using
various channels such as contact emails / forms, LinkedIn outreach, and
personal connections. Strong applications were invited to interviews and
eventually to participate in a paid 3-day work trial. Compensation happened
according to the respective rates of the consultancy. 

The work trial consisted of a three day 2 FTE implementation of an
adapted unlearning method\footnote{In short, we asked consultancies to
implement Qian et al.\ [12] in the RTT codebase by
Deeb et al.\ [13].} (research code repository in the Supplements).
Each team was provided with a Slack channel to discuss progress on the
task and point out any upcoming questions / issues. The final
submission, including git repository, end-of-work-trial report, and
group chat history was then scored on a 0 to 3 scale by two independent
ML research staff on a weighted metric, including Technical
Correctness (40\,\%), Experimental Rigor \& Reproducibility (25\,\%),
Code Quality \& Maintainability (15\,\%), Communication \& Insight
(15\,\%), and Bonus Innovation / Extras (up to 10\%) (see Supplements
for the evaluation metrics). Final scores were averaged. As a control we
asked a technical staff member from METR to let modern AI coding agents
(Claude Code, GPT-5) execute the task as well.

\clearpage
\section{References}

\begin{enumerate}
\item Alphabet Investor Relations. Google 10-K Annual Report 2024. \url{https://abc.xyz/assets/77/51/9841ad5c4fbe85b4440c47a4df8d/goog-10-k-2024.pdf}

\item Microsoft Investor Relations. FY 2024 Q4 Earnings. \url{https://www.microsoft.com/en-us/investor/events/fy-2024/earnings-fy-2024-q4}

\item Meta Investor Relations. Third Quarter 2024 Results. \url{https://investor.atmeta.com/investor-news/press-release-details/2024/Meta-Reports-Third-Quarter-2024-Results/default.aspx}

\item About Amazon. Amazon AWS Anthropic AI. \url{https://www.aboutamazon.com/news/company-news/amazon-aws-anthropic-ai}

\item ML Alignment Theory Scholars (MATS) Program. \url{https://www.matsprogram.org/}

\item Accenture Newsroom. Accenture to invest \$3 billion in AI. \url{https://newsroom.accenture.com/news/2023/accenture-to-invest-3-billion-in-ai-to-accelerate-clients-reinvention}

\item Capgemini. Augmented Engineering Offerings Powered by Gen AI. \url{https://www.capgemini.com/news/press-releases/capgemini-announces-augmented-engineering-offerings-powered-by-gen-ai/}

\item LinkedIn Statistics for Professionals. \url{https://www.linkedin.com/pulse/50-linkedin-statistics-every-professional-should-ti9ue/}

\item Fatehkia M, et al. Mapping socioeconomic indicators using social media advertising data. \textit{EPJ Data Science}. 2021. \url{https://epjdatascience.springeropen.com/articles/10.1140/epjds/s13688-021-00294-7}

\item Barkay N, et al. Coverage properties of LinkedIn data. \textit{Journal of Survey Statistics and Methodology}. 2024;12(5):1200. \url{https://academic.oup.com/jssam/article/12/5/1200/7728213}

\item World Bank Group. LinkedIn Data Insights: Jobs, Skills and Migration Trends. \url{https://documents1.worldbank.org/curated/en/827991542143093021/pdf/World-Bank-Group-LinkedIn-Data-Insights-Jobs-Skills-and-Migration-Trends-Methodology-and-Validation-Results.pdf}

\item Qian J, et al. 2025. arXiv:2505.09500. \url{https://arxiv.org/abs/2505.09500}

\item Deeb S, et al. 2024. arXiv:2410.08827. \url{https://arxiv.org/html/2410.08827v1}
\end{enumerate}

\clearpage
\appendix
\renewcommand{\thefigure}{S\arabic{figure}}
\setcounter{figure}{0}
\renewcommand{\thetable}{S\arabic{table}}
\setcounter{table}{0}
\section{Supplements}\label{h.tdxac6bhdlxa}

\subsection{Repository}\label{h.frq7f14a5lg6}

The following repository includes all analysis and work trial code. It
only includes partial information on the exported profiles.

\url{https://github.com/MxSchons-GmbH/consultancy-ml-research-talent-estimates}

\subsection{Extended Flow Chart}\label{h.qa2z0fubsm3c}

\begin{figure}[htbp]
\centering
\pandocbounded{\includegraphics[keepaspectratio]{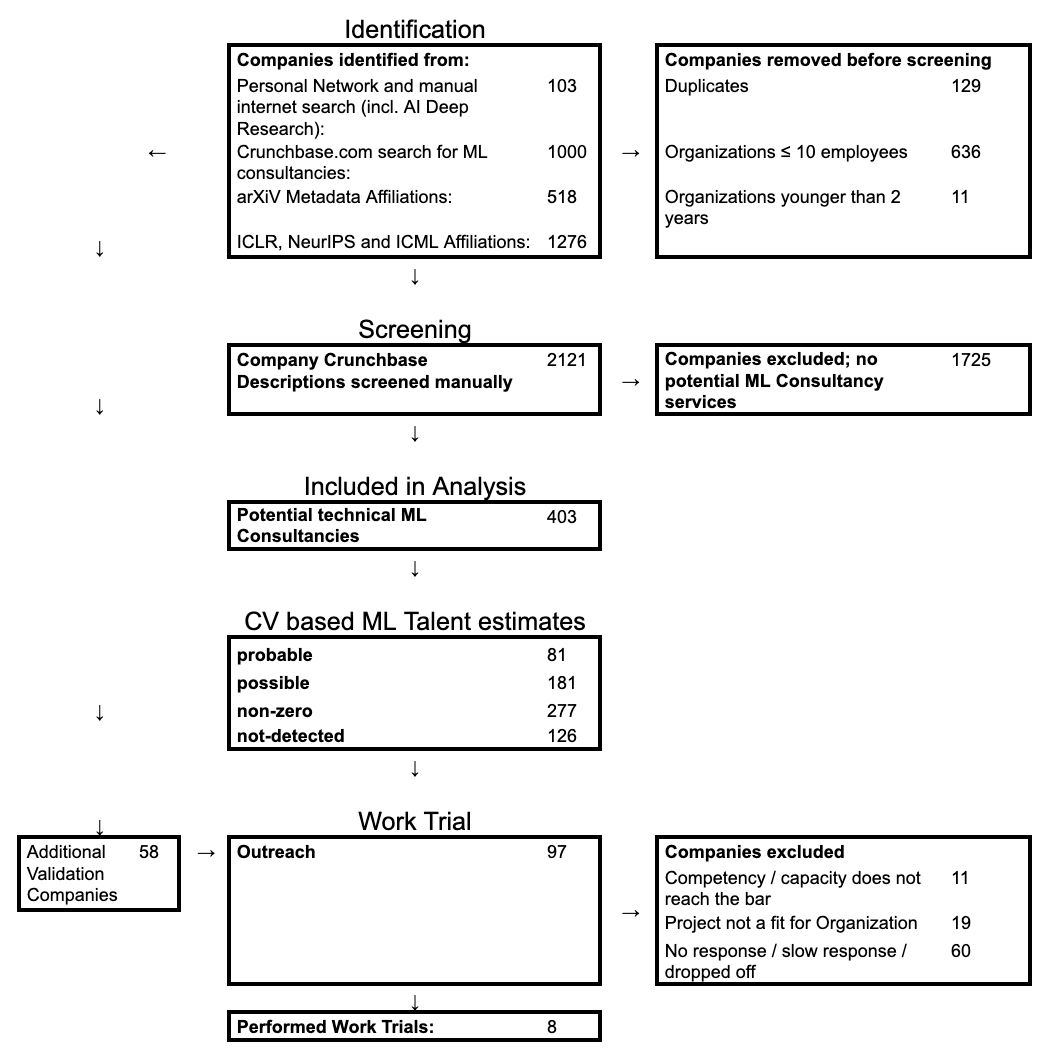}}
\caption{Extended flow chart of the systematic search and estimation pipeline}
\label{fig:extended-flowchart}
\end{figure}

\subsection{Search strategies}\label{h.tjifkul59853}

\begin{enumerate}
\tightlist
\item
  {Team's contacts and its networks recommendations}
\item
  {LLM deep research based}
\item
  {Unstructured Web Search with two independent RAs}
\item
  {Research \& publication venues: arXiv Metadata (see code repository
  for details)}
\item
  {Conferences \& sponsor/exhibitor lists: ICLR, ICML and NeurIPS 2019
  onwards (see code repository for details)}
\item
  {Crunchbase Pro + keyword filters}
\end{enumerate}

\medskip
\noindent
\begin{minipage}{\linewidth}
\small\sffamily\bfseries Crunchbase Pro Query Filters\par
\vspace{4pt}
\scriptsize\rmfamily\mdseries
\setlength{\fboxsep}{8pt}
\noindent\fbox{\parbox{\dimexpr\linewidth-2\fboxsep-2\fboxrule}{%
\textbf{INDUSTRY:} Includes (Outsourcing, Consulting)\par\smallskip
\textbf{FOUNDED DATE:} Before (01/01/2022)\par\smallskip
\textbf{FULL DESCRIPTION:} Includes (artificial intelligence, machine learning,
deep learning, ai, ml, LLM, large language model, reinforcement
learning)\par\smallskip
\textbf{DESCRIPTION:} Includes (artificial intelligence, machine learning, deep
learning, ai, ml, LLM, large language model, reinforcement learning)\par\smallskip
\textbf{OPERATING STATUS:} Does not include (Closed)\par\smallskip
\textbf{FULL DESCRIPTION:} Does not include (SaaS, Cloud-based analytics, Cloud infrastructure, Biodegradability testing, Enterprise software, Data center, Data capture, Document management, Semiconductor, Venture studio, Credit bureau reporting, Data labeling, Training data curation, Business growth consulting, Blockchain, Geospatial database, Incubator, Acquiring target audiences, Pyrotechnic products, Proprietary, self-service, Quantum software, software framework, Business development and communications, Leadership development, Sales enablement, Graphic Design Studio, Cloud Provider, GenAI Platform, PaaS, IaaS, Low-code application, Exoplanet discovery, ARM industry, Credit risk, Commercialization, subsurface environments, Brand intelligence, Customer relationship, Content creation, Podcast, Quantitative finance, Mobile app, Conference, Video analytics, Servicetitan, Call center, ACMI, Electric fan, Investment advisory, Market intelligence, Low code, web mobile, real estate, web development, financial services, cyber security, digital marketing, robotic process automation, robotic process, ui ux, mobile application, mobile development, ar vr, ux design, ux ui, mobile application development, development mobile, virtual reality, ios android, search engine, mobile apps, ui design, ui ux design, neo technologies, maktoum neo technologies, ux ui design, web development mobile, ai blockchain, website development, web mobile development, web mobile application development, web mobile application, blockchain development, blockchain ai, phone calls, sales service, web application, web scraping, web services, telefonica open, intelligence blockchain, iot artificial, iot data, consulting web, content marketing, marketing services, marketing solutions, search engine optimization, retail commerce, marketing sales, web applications, web design, web app, supply chain, management consulting, internet things, design development, social media, strategic consulting, market research, iot, cloud, strategy consulting, sales marketing, data security, legal services, media scope, oil gas, law firms, injury illness, analytics big data, E-commerce, Subscription, Data labelling, Ai-powered platform, Virtual assistant, Global brands, Intellectual property solutions, Industrial engineering, Intellectual property solutions, Intellectual property, Geochemical, Lead conversion, Workforce management, Personal learning products, Design-led, Translation agency, Online robot lawyer, Business process architecture, Contact center, Salesforce, Ophthalmology, Cultural tourism, Emotion analytics, EEG, Jobs marketplace, MIPS services, Product storytelling, Smart Signals, Revenue-raising, Licensing, online learning, Moodle-based, staffing and recruitment, AI-powered growth partner, AI-powered language technology, AI engine, Wellbeing support, Ambient intelligence, Marketing Mix, Transaction flows, Human resources technology, Marketplace, asset health, Smart recruitment, CRO, Annotation services, Safety solution strategy, Asset of value, Innovation intelligence, Competitive intelligence, High quality courses)%
}}
\end{minipage}
\medskip

We did not explicitly search through startup \& accelerator programs,
nor make any big public announcements in mailing lists or run ads. Other
data streams we either did not explore or were not satisfied with for
getting data on companies: GitHub, Kaggle teams leaderboard, Gartner
Market Guide, Forrester Wave, PitchBook, CapitalIQ, CB Insights, G2,
Clutch, GoodFirms. The search was only in the English language. We
investigated government repositories, but it wasn't helpful as keywords
like machine learning weren't available.

\subsection{Specs for identification}\label{h.q8bq2s7i33aq}

Keyword extraction was carried out in Google Colab with KeyBERT, which
ranks candidate 1- to 4-gram phrases by computing cosine similarity
between their TF-IDF-weighted representations and sentence-level
embeddings generated by the Salesforce/SFR-Embedding-Mistral model from
SentenceTransformers running on an NVIDIA A100 GPU via PyTorch. 

\subsection{Sensitivity / Specificity across models}\label{h.x3pe4fqvr14w}

True / False Positives and Negatives, Sensitivity /
Specificity and F1 value of various LinkedIn Search methods

{\pandocbounded{\includegraphics[keepaspectratio]{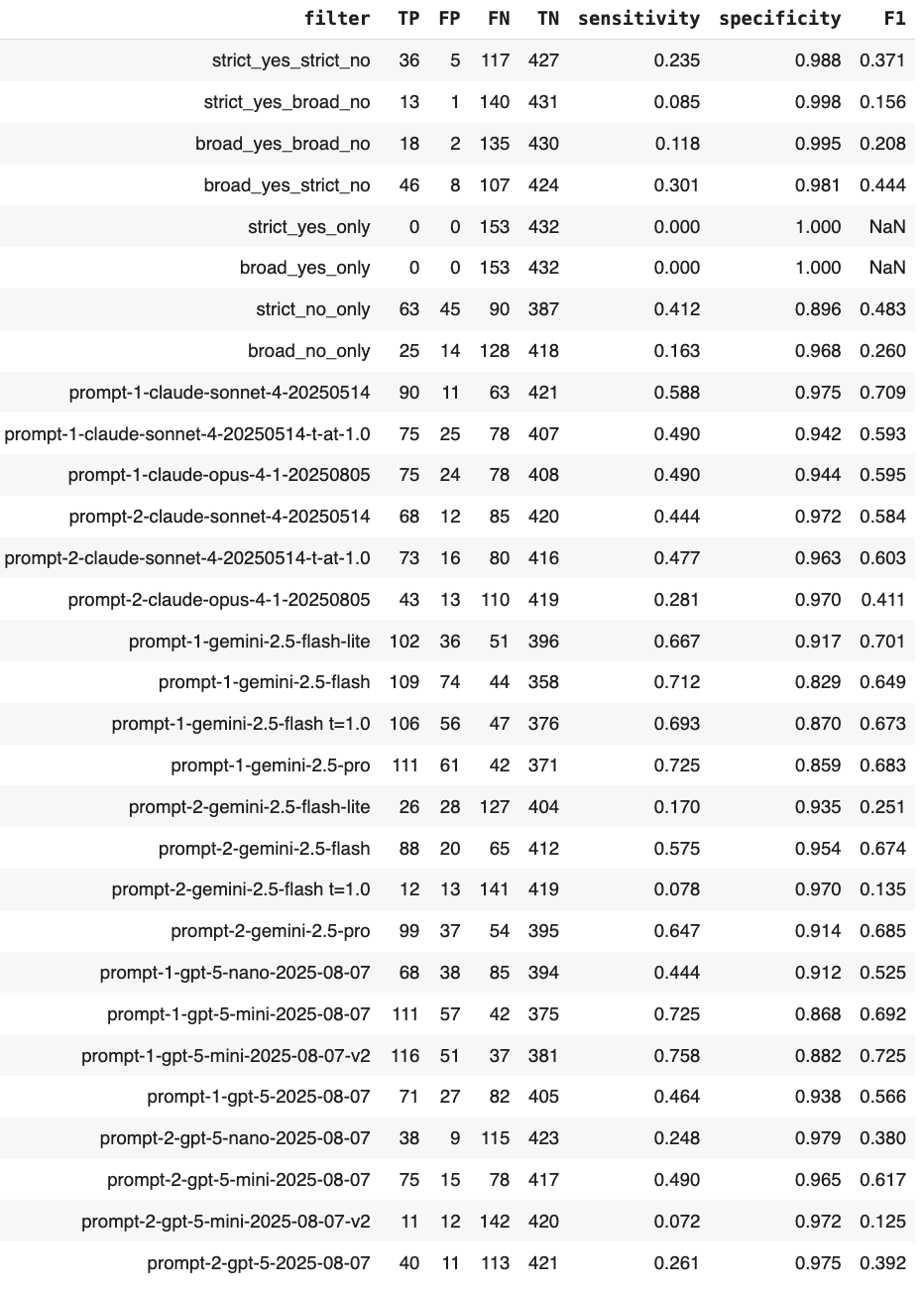}}}

\scriptsize
\begin{longtable}[]{@{}p{1.3in}p{0.9in}p{0.9in}p{0.9in}@{}}
\toprule\noalign{}
\endhead
\bottomrule\noalign{}
\endlastfoot
{\textbf{Annotator}} & {\textbf{Sensitivity}} & {\textbf{Specificity}} & {\textbf{Accuracy}} \\
{correlated\_probit} & {0.791} & {0.926} & {0.891} \\
\end{longtable}
\normalsize
\renewcommand{\arraystretch}{1.2}

\subsection{Validation Dataset}\label{h.i62j3fl62f28}

Available in code repository.

Profiles included the following companies:

Small-scale technical AI Alignment: Palisade Research, CHAI, FAR.AI,
Apollo Research, Goodfire and Transluce. Large-scale technical AI
Alignment: Google DeepMind, OpenAI, and Anthropic

IT Consultancies: Accenture, HCL Technologies, Ernst \& Young (EY),
Infosys, Wipro, PricewaterhouseCoopers (PwC), Cognizant, Tata
Consultancy Services, Deloitte, Capgemini, Bain, BCG X

\subsection{Validation: AI and non-AI companies}\label{h.p0t7lwao63c6}

As an additional validation step, we applied the selected estimation
methods to companies we knew to have high technical ML Research Talent counts and
to organizations who certainly don't. The results are presented in the
Supplements. Note that some companies were not evaluated with LLMs due
to their size and associated costs.

\begin{figure}[htbp]
\centering
\pandocbounded{\includegraphics[keepaspectratio]{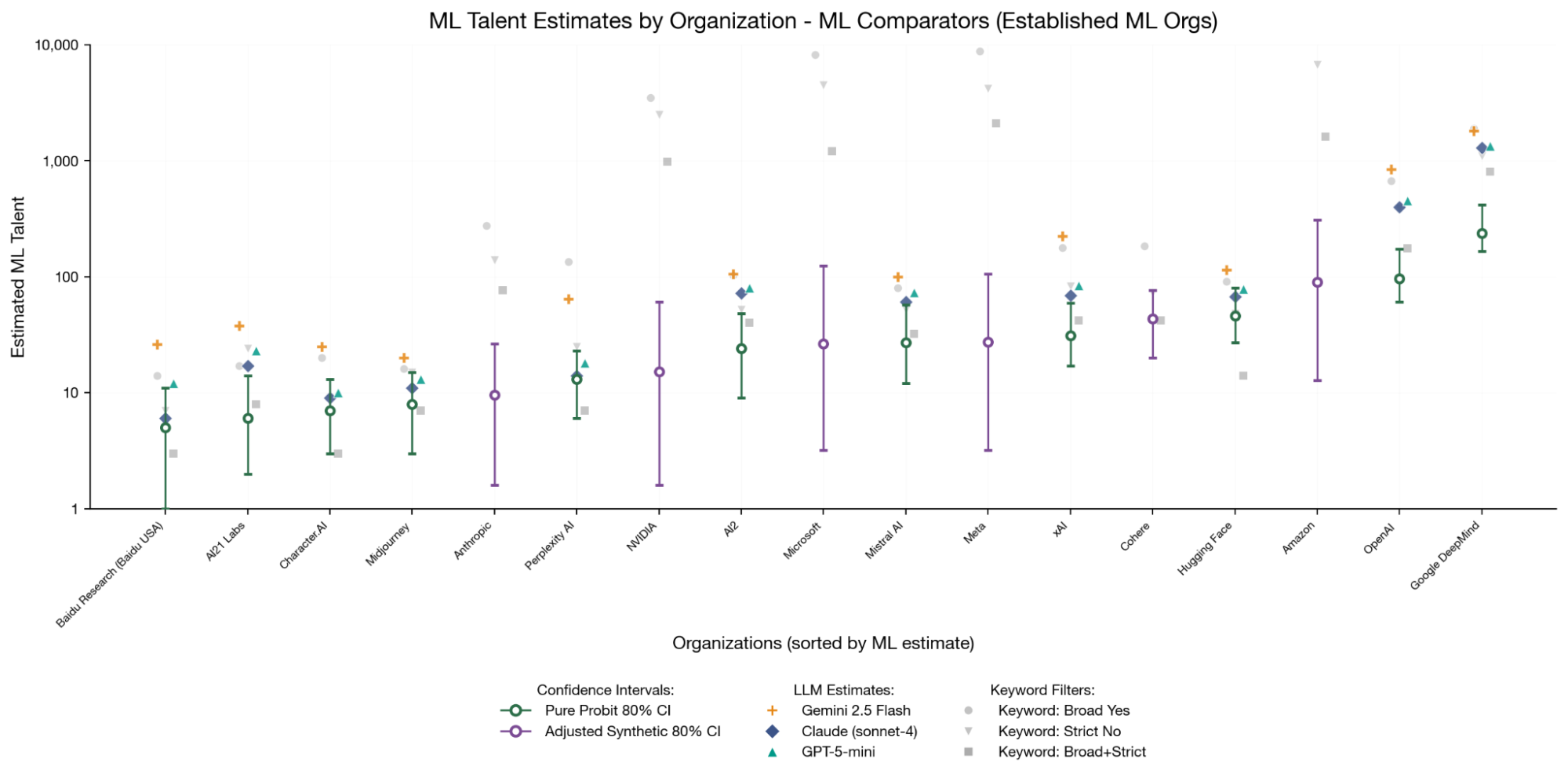}}
\pandocbounded{\includegraphics[keepaspectratio]{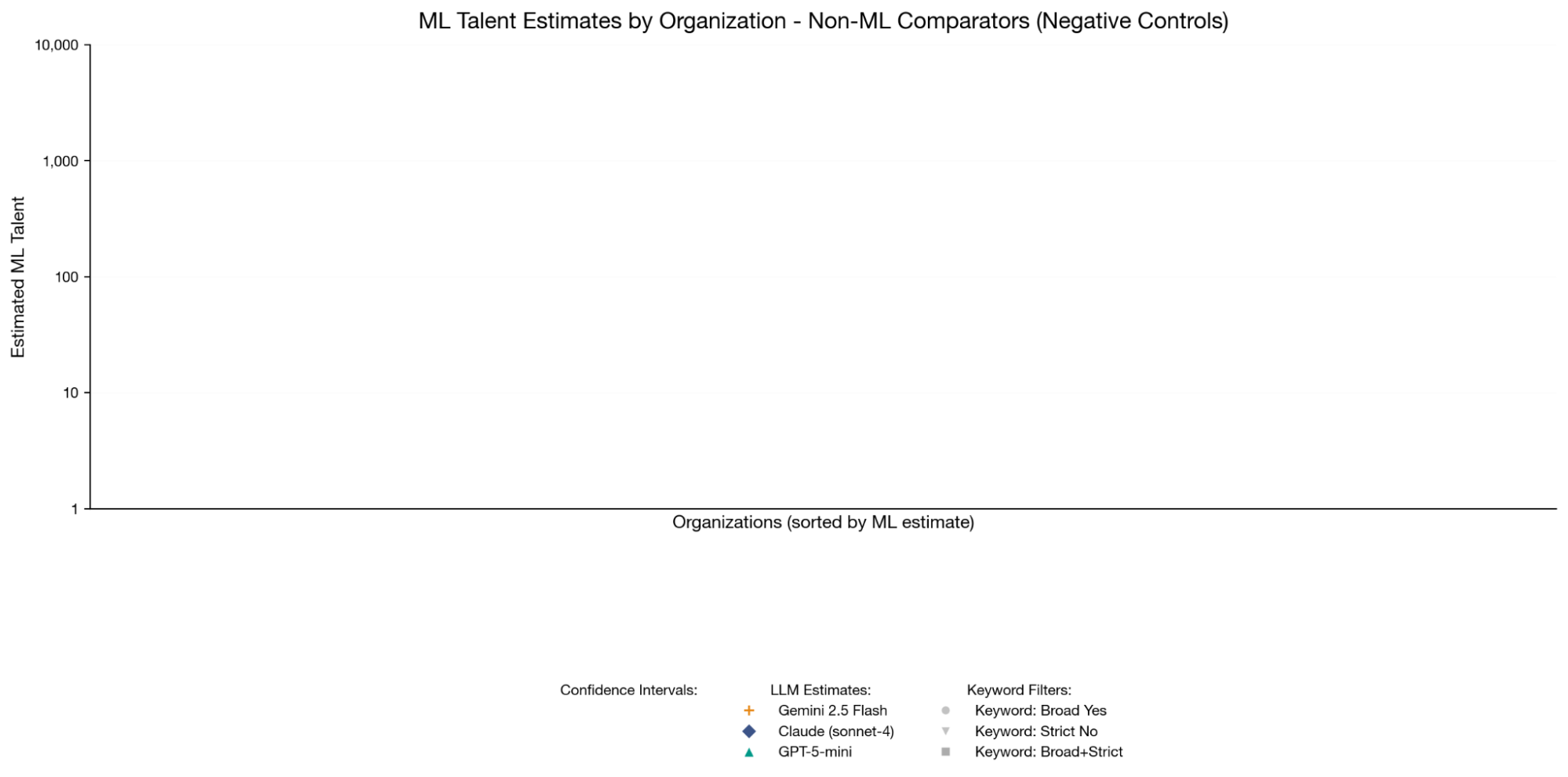}}
\caption{Validation: AI and non-AI company estimates}
\label{fig:validation}
\end{figure}

\clearpage
\scriptsize
\begin{longtable}[]{@{}
  >{\raggedright\arraybackslash}p{(\linewidth - 14\tabcolsep) * \real{0.14}}
  >{\raggedright\arraybackslash}p{(\linewidth - 14\tabcolsep) * \real{0.08}}
  >{\raggedright\arraybackslash}p{(\linewidth - 14\tabcolsep) * \real{0.10}}
  >{\raggedright\arraybackslash}p{(\linewidth - 14\tabcolsep) * \real{0.10}}
  >{\raggedright\arraybackslash}p{(\linewidth - 14\tabcolsep) * \real{0.22}}
  >{\raggedright\arraybackslash}p{(\linewidth - 14\tabcolsep) * \real{0.14}}
  >{\raggedright\arraybackslash}p{(\linewidth - 14\tabcolsep) * \real{0.12}}
  >{\raggedright\arraybackslash}p{(\linewidth - 14\tabcolsep) * \real{0.10}}@{}}
\toprule\noalign{}
\endhead
\bottomrule\noalign{}
\endlastfoot
{\textbf{Company Name}} & {\textbf{Founded}} & {\textbf{Country}} & {\textbf{Total Staff (LinkedIn)}} &
{\textbf{Individual Estimates [broad+strict, strict, broad, claude, gpt5,
gemini]}} & {\textbf{ML Research Talent q50 (q10 -- q90)}} & {\textbf{ML \% of Total}} &
{\textbf{Category}} \\
{Google DeepMind} & {1970} & {Unknown} & {6 500} & {[801, 1100, 1900,
1285, 1330, 1803]} & {237 (166 - 414)} & {3.65\% (2.55\% - 6.38\%)} &
{Probable} \\
{OpenAI } & {1970} & {Unknown} & {6 500} & {[175, 391, 671, 397, 454,
845]} & {96 (61 - 173)} & {1.48\% (0.94\% - 2.66\%)} & {Probable} \\
{Amazon} & {1970} & {Unknown} & {540 000} & {[1600, 6700, 11000, -, -,
-]} & {89 (12 - 308) *} & {0.02\% (0.00\% - 0.06\%)} & {Probable} \\
{Hugging Face} & {1970} & {Unknown} & {603} & {[14, 48, 91, 67, 78,
114]} & {46 (27 - 80)} & {7.63\% (4.48\% - 13.27\%)} & {Probable} \\
{Cohere} & {1970} & {Unknown} & {810} & {[42, 73, 183, -, -, -]} &
{43 (20 - 76) *} & {5.33\% (2.47\% - 9.38\%)} & {Probable} \\
{xAI} & {1970} & {Unknown} & {3 000} & {[42, 83, 177, 69, 84, 223]}
& {31 (17 - 59)} & {1.03\% (0.57\% - 1.97\%)} & {Probable} \\
{Mistral AI} & {1970} & {Unknown} & {440} & {[32, 52, 80, 61, 73,
100]} & {27 (12 - 57)} & {6.14\% (2.73\% - 12.95\%)} & {Probable} \\
{Meta} & {1970} & {Unknown} & {70 000} & {[2100, 4200, 8800, -, -,
-]} & {27 (3 - 105) *} & {0.04\% (0.00\% - 0.15\%)} & {Probable} \\
{Microsoft} & {1970} & {Unknown} & {22 000} & {[1200, 4500, 8200, -,
-, -]} & {26 (3 - 123) *} & {0.12\% (0.01\% - 0.56\%)} & {Probable} \\
{AI2} & {1970} & {Unknown} & {313} & {[40, 52, 103, 72, 80, 106]} &
{24 (9 - 48)} & {7.67\% (2.88\% - 15.34\%)} & {Probable} \\
{NVIDIA} & {1970} & {Unknown} & {43 000} & {[974, 2500, 3500, -, -,
-]} & {15 (1 - 60) *} & {0.04\% (0.00\% - 0.14\%)} & {Probable} \\
{Perplexity AI} & {1970} & {Unknown} & {1 000} & {[7, 25, 135, 14, 18,
64]} & {13 (6 - 23)} & {1.30\% (0.60\% - 2.30\%)} & {Probable} \\
{Anthropic} & {1970} & {Unknown} & {2 000} & {[76, 140, 276, -, -,
-]} & {9 (1 - 26) *}
& {0.48\% (0.08\% - 1.32\%)} & {Probable} \\
{Midjourney} & {1970} & {Unknown} & {171} & {[7, 15, 16, 11, 13,
20]} & {8 (3 - 15)} & {4.68\% (1.75\% - 8.77\%)} & {Probable} \\
{Character.AI} & {1970} & {Unknown} & {207} & {[3, 9, 20, 9, 10,
25]} & {7 (3 - 13)} & {3.38\% (1.45\% - 6.28\%)} & {Probable} \\
{AI21 Labs} & {1970} & {Unknown} & {250} & {[8, 24, 17, 17, 23, 38]}
& {6 (2 - 14)} & {2.40\% (0.80\% - 5.60\%)} & {Probable} \\
{Baidu Research (Baidu USA)} & {1970} & {Unknown} & {238} & {[3, 7,
14, 6, 12, 26]} & {5 (1 - 11)} & {2.10\% (0.42\% - 4.62\%)} &
{Probable} \\
{LawZero} & {1970} & {Unknown} & {11} & {[1, 2, 4, 4, 5, 5]} & {2 (0
- 5)} & {18.18\% (0.00\% - 45.45\%)} & {Possible} \\
{Thinking Machines Lab} & {1970} & {Unknown} & {67} & {[12, 15, 22, -,
-, -]} & {0 (0 - 0) *} & {0.00\% (0.00\% - 0.00\%)} & {Not
Detected} \\
{Mila-Quebec Artificial Intelligence Institute} & {1970} & {Unknown} &
{872} & {[130, 163, 424, -, -, -]} & {0 (0 - 0) *} & {0.00\% (0.00\%
- 0.00\%)} & {Not Detected} \\
\end{longtable}
\normalsize
\renewcommand{\arraystretch}{1.2}

\clearpage
\scriptsize
\begin{longtable}[]{@{}
  >{\raggedright\arraybackslash}p{(\linewidth - 14\tabcolsep) * \real{0.14}}
  >{\raggedright\arraybackslash}p{(\linewidth - 14\tabcolsep) * \real{0.08}}
  >{\raggedright\arraybackslash}p{(\linewidth - 14\tabcolsep) * \real{0.10}}
  >{\raggedright\arraybackslash}p{(\linewidth - 14\tabcolsep) * \real{0.10}}
  >{\raggedright\arraybackslash}p{(\linewidth - 14\tabcolsep) * \real{0.22}}
  >{\raggedright\arraybackslash}p{(\linewidth - 14\tabcolsep) * \real{0.14}}
  >{\raggedright\arraybackslash}p{(\linewidth - 14\tabcolsep) * \real{0.12}}
  >{\raggedright\arraybackslash}p{(\linewidth - 14\tabcolsep) * \real{0.10}}@{}}
\toprule\noalign{}
\endhead
\bottomrule\noalign{}
\endlastfoot
{\textbf{Company Name}} & {\textbf{Founded}} & {\textbf{Country}} & {\textbf{Total Staff (LinkedIn)}} &
{\textbf{Individual Estimates [broad+strict, strict, broad, claude, gpt5,
gemini]}} & {\textbf{ML Research Talent q50 (q10 -- q90)}} & {\textbf{ML \% of Total}} &
{\textbf{Category}} \\
{Burberry} & {1970} & {Unknown} & {8 500} & {[0, 2, 5, 1, 1, 2]} &
{1 (0 - 3)} & {0.01\% (0.00\% - 0.04\%)} & {Possible} \\
{Sherwin-Williams} & {1970} & {Unknown} & {44 000} & {[2, 29, 28, 1,
2, 6]} & {1 (0 - 3)} & {0.00\% (0.00\% - 0.01\%)} & {Possible} \\
{The Coca-Cola Company} & {1970} & {Unknown} & {60 000} & {[2, 31, 68,
3, 9, 30]} & {1 (0 - 5)} & {0.00\% (0.00\% - 0.01\%)} & {Possible} \\
{The North Face} & {1970} & {Unknown} & {4 500} & {[0, 0, 2, 0, 0,
0]} & {0 (0 - 0)} & {0.00\% (0.00\% - 0.00\%)} & {Not Detected} \\
{Desigual} & {1970} & {Unknown} & {2 500} & {[0, 0, 0, 0, 0, 1]} &
{0 (0 - 0)} & {0.00\% (0.00\% - 0.00\%)} & {Not Detected} \\
{Jotun} & {1970} & {Unknown} & {6 000} & {[0, 1, 5, 0, 0, 2]} & {0
(0 - 0)} & {0.00\% (0.00\% - 0.00\%)} & {Not Detected} \\
{Sierra Nevada Brewing Co.} & {1970} & {Unknown} & {769} & {[0, 0, 0,
0, 0, 0]} & {0 (0 - 0)} & {0.00\% (0.00\% - 0.00\%)} & {Not
Detected} \\
{Penguin Random House} & {1970} & {Unknown} & {8 000} & {[1, 9, 18, 2,
2, 7]} & {0 (0 - 2)} & {0.00\% (0.00\% - 0.03\%)} & {Non-zero} \\
{HarperCollins Publishers} & {1970} & {Unknown} & {4 000} & {[0, 8, 7,
0, 0, 1]} & {0 (0 - 2)} & {0.00\% (0.00\% - 0.05\%)} & {Non-zero} \\
{Patagonia} & {1970} & {Unknown} & {3 500} & {[0, 1, 0, 0, 0, 0]} &
{0 (0 - 0)} & {0.00\% (0.00\% - 0.00\%)} & {Not Detected} \\
{Crocs, Inc.} & {1970} & {Unknown} & {5 500} & {[0, 1, 3, 0, 0, 0]}
& {0 (0 - 0)} & {0.00\% (0.00\% - 0.00\%)} & {Not Detected} \\
{The British Museum} & {1970} & {Unknown} & {1 000} & {[0, 1, 0, 0, 0,
0]} & {0 (0 - 0)} & {0.00\% (0.00\% - 0.00\%)} & {Not Detected} \\
{Museo Nacional del Prado} & {1970} & {Unknown} & {296} & {[0, 0, 0,
0, 0, 0]} & {0 (0 - 0)} & {0.00\% (0.00\% - 0.00\%)} & {Not
Detected} \\
{London Symphony Orchestra} & {1970} & {Unknown} & {232} & {[0, 0, 0,
0, 0, 0]} & {0 (0 - 0)} & {0.00\% (0.00\% - 0.00\%)} & {Not
Detected} \\
{New York Philharmonic} & {1970} & {Unknown} & {288} & {[0, 0, 0, 0,
0, 0]} & {0 (0 - 0)} & {0.00\% (0.00\% - 0.00\%)} & {Not Detected} \\
{Hydro Flask} & {1970} & {Unknown} & {135} & {[0, 1, 0, 0, 0, 0]} &
{0 (0 - 0)} & {0.00\% (0.00\% - 0.00\%)} & {Not Detected} \\
{YETI} & {1970} & {Unknown} & {2 000} & {[0, 0, 5, 1, 0, 1]} & {0 (0
- 0)} & {0.00\% (0.00\% - 0.00\%)} & {Not Detected} \\
{LVMH} & {1970} & {Unknown} & {7 000} & {[0, 2, 17, 1, 5, 12]} & {0
(0 - 0)} & {0.00\% (0.00\% - 0.00\%)} & {Not Detected} \\
\end{longtable}
\normalsize
\renewcommand{\arraystretch}{1.2}

\clearpage
\subsection{Work Trial Repository and Evaluation Metric}\label{h.9xegcc3ajgl8}

\url{https://github.com/MxSchons-GmbH/consultancy-ml-research-talent-estimates}

1. Technical Correctness (Weight\,=\,40\,\%)

\scriptsize
\begin{longtable}[]{@{}p{0.9in}p{4.3in}@{}}
\toprule\noalign{}
\endhead
\bottomrule\noalign{}
\endlastfoot
{\textbf{Score}} & {\textbf{Description}} \\
{3 --\,Excellent} & {All required artefacts present; metrics reported
for }{Forget\,T, Validation\,V, Retain\,R }{before and after RTT}{;
Recovery‑Rate stated and $\leq\,1.05$ (i.e. $\leq\,105\,\%$ of original
performance); code implements correct data splits and loss functions
described in RTT (§4.3)\,.} \\
{2 --\,Minor flaws} & {Runs finish and metrics exist, but one of: (i)
wrong split sizes, (ii) Recovery‑Rate omitted, (iii) retain‑drop
exceeds stated 5\,/\,10\,/\,30\,\% boundary without explanation\,.} \\
{1 --\,Major flaws} & {Pipeline executes but results conflict with spec
(e.g., forget accuracy }{increases after unlearning\,); or RTT
fine‑tunes the }{original not the unlearned model} \\
{0 --\,Invalid} & {Script crashes, wrong target task, or any hand‑edited
prediction/metric file detected (MLE‑Bench red‑flag).} \\
\end{longtable}
\normalsize
\renewcommand{\arraystretch}{1.2}

LLM actions: parse main.py AST for data‑loader arguments; diff reported
numbers against wandb\_run‑*.json; compute Recovery‑Rate automatically.

\begin{center}\rule{0.5\linewidth}{0.5pt}\end{center}

2. Experimental Rigor \& Reproducibility (25\,\%)

\scriptsize
\begin{longtable}[]{@{}p{0.9in}p{4.3in}@{}}
\toprule\noalign{}
\endhead
\bottomrule\noalign{}
\endlastfoot
{\textbf{Score}} & {\textbf{Description}} \\
{3} & {Clear baseline run, $\geq$\,1 hyper‑parameter sweep for both baseline
\& method; fixed seeds (torch.manual\_seed or Hydra seed=); compute
budgets (epochs, GPU type) stated; W\&B run IDs and Git commit SHA
logged.} \\
{2} & {Baseline exists or seeds fixed, but hyper‑sweep partial
(\textless3 values) or compute not recorded.} \\
{1} & {Only a single run; no baseline; ad‑hoc HP search (``time
constraints'')\,.} \\
{0} & {No experimental details or logs recoverable.} \\
\end{longtable}
\normalsize
\renewcommand{\arraystretch}{1.2}

\begin{center}\rule{0.5\linewidth}{0.5pt}\end{center}

3. Code Quality \& Maintainability (15\,\%)

LLM parses the lone entry‑point plus config.

\scriptsize
\begin{longtable}[]{@{}p{0.9in}p{4.3in}@{}}
\toprule\noalign{}
\endhead
\bottomrule\noalign{}
\endlastfoot
{\textbf{Score}} & {\textbf{Description}} \\
{3} & {PEP8‑clean; functions/classes \textgreater\,70\,\% documented;
typing used; cyclomatic complexity \textless\,15; configs externalised
(Hydra / YAML).} \\
{2} & {Modular but sparse docs or missing types; minor duplication; lint
passes with $\leq\,10$\,warnings.} \\
{1} & {Monolithic script, hard‑coded paths/HPs,
\textgreater\,30\,warnings, acknowledged need for refactor.} \\
{0} & {Won't import, mixed tabs/spaces, or contains credentials.} \\
\end{longtable}
\normalsize
\renewcommand{\arraystretch}{1.2}

\begin{center}\rule{0.5\linewidth}{0.5pt}\end{center}

4. Communication \& Insight (15\,\%)

\scriptsize
\begin{longtable}[]{@{}p{0.9in}p{4.3in}@{}}
\toprule\noalign{}
\endhead
\bottomrule\noalign{}
\endlastfoot
{\textbf{Score}} & {\textbf{Report Attributes}} \\
{3} & {Executive summary, method diagram/figure, interprets failure
modes, lists next‑steps; cites RTT correctly; $\leq\,10$ pages; professional
tone.} \\
{2} & {Complete narrative but thin reflection (e.g., ``time constraints''
repeated).} \\
{1} & {Mostly raw W\&B screenshots; prose \textless\,1\,page; vague
claims (``results look good'').} \\
{0} & {Missing or incomprehensible.} \\
\end{longtable}
\normalsize
\renewcommand{\arraystretch}{1.2}

\begin{center}\rule{0.5\linewidth}{0.5pt}\end{center}

5. Bonus Innovation / Extras (+0‑10\,ppt)

Award in 2‑point increments for: ablation studies, alternative
unlearning algorithms, automated sweep scripts, CI tests, dashboards,
compute‑matched comparisons, etc. (Cap bonus at\,+10\,ppt so core
metrics dominate.)

\begin{center}\rule{0.5\linewidth}{0.5pt}\end{center}

6. Aggregation

Final \% = 0.40$\cdot$A1 + 0.25$\cdot$A2 + 0.15$\cdot$A3 + 0.15$\cdot$A4 + Bonus

Recommendation: \textgreater{} 70\%\\
Conditional Recommendation: $\geq$ 50\%

No Recommendation: \textless{} 50\%

\begin{center}\rule{0.5\linewidth}{0.5pt}\end{center}

\subsection{Specification of company profiles and talent}\label{h.8d4cg3n590m}

Applicants who passed a minimum bar of subject matter expertise (e.g.
at least one strong ML researcher or past projects that demonstrated
substantive research abilities) were invited to an interview. If
applicants didn't meet the bar they were offered the opportunity to
provide different team members. Interviews served for further validation
of subject matter expertise (``share the two to three most complex
machine learning projects you worked on in the past 6-12 months''),
internal scale (``How many employees do you have who could tackle tasks
such as training GPT-2 from scratch''), research approach (``Would you
want to drive the research internally or do you need external
guidance''), and providing additional context for the IT consultancies.

\scriptsize
\begin{longtable}[]{@{}
  >{\raggedright\arraybackslash}p{(\linewidth - 0\tabcolsep) * \real{1.0000}}@{}}
\toprule\noalign{}
\endhead
\bottomrule\noalign{}
\endlastfoot
\begin{minipage}[t]{\linewidth}\raggedright
{\textbf{Ideal Company Profile}}

Here are some ideal attributes of organizations we are looking for (but
not necessarily dealbreakers if a vendor does not satisfy 1 of these
criteria).

Our goal is to test if potential consultancies in industry can
contribute meaningfully to research in AI safety. As such, we have
focused on the following attributes:

\begin{enumerate}
\tightlist
\item
  Consultancies that have existed or operated actively with client work
  for more than 2 years as an indicator of organizational execution
\item
  Organizations that have upwards of 15 staff (long-term contracts or
  permanent staff), or at least have demonstrated ability to manage
  increasing staff overhead successfully over time
\item
  Organizations that have enough liquidity to complete the project,
  ideally demonstrating they have the necessary resources
  to complete the 2-month project and manage their financial risks
\item
  Consultancies that can demonstrate that their project portfolio has a
  good amount of diversity (or at least that the ML work is not only
  related to AI safety research)
\end{enumerate}

Unfortunately, we have decided to not prioritize contractor network
groups (where team members are usually on a project-basis and are not
long-term or permanent staff members) as our aim is to test
consultancies that have some longer-term cohesion.

Staff Roles Required

\subsubsection{Researcher (at least 1)}\label{h.vn9l98u676of}

Focuses on conceptual understanding, creative problem-solving, and
theoretical knowledge.

Responsibilities

\begin{itemize}
\tightlist
\item
  {Attempt to understand the research problem deeply.}
\item
  {Explore and articulate ideas and the approaches for creating the
  DMUM.}
\item
  {Create detailed method specifications for the DMUM.}
\item
  {Provide reports on progress, challenges, solutions, and assumptions.}
\item
  {Shows excitement about potentially publishing research findings.}
\end{itemize}

Desired traits

\begin{itemize}
\tightlist
\item
  {Expertise in relevant academic fields, potentially evidenced by
  papers written with company affiliation or authorship. PhD is ideal
  but not required.}
\item
  {Experience with public outputs (e.g., blogs, GitHub repos). Some
  engagement with relevant online communities on the research problem
  ideal but not required}
\item
  {Tenacity in tackling challenges and asking pertinent clarifying
  questions.}
\item
  {Comfortable working with Large Language Models (LLMs).}
\item
  {Adept at identifying and isolating root causes of unexpected research
  outcomes or model behaviors.}
\end{itemize}

\subsubsection{ML Researcher (at least 1)}\label{h.dp0ghuwpnjh}

Focuses on the practical implementation, code integration, and ensuring
functional output.

Responsibilities

\begin{itemize}
\tightlist
\item
  {Understand the existing codebase(s), modify or add method logic, and
  create necessary configurations}
\item
  {Implement methods according to specifications (ensure robust loss
  calculation and handling of the model's internal
  mechanisms)}
\item
  {Ensures code runs without crashing and produces expected results,
  even if preliminary}
\item
  {Provide clear and timely progress reports (e.g., Loom/text reports)
  detailing activities, challenges, and solutions.}
\end{itemize}

Desired Background/Traits:

\begin{itemize}
\tightlist
\item
  {Comfortable building implementations directly from specifications}
\item
  {Comfortable with manipulating various data formats and handling their
  integration into the codebase}
\item
  {Comfortable with distributed runs and
  \href{https://arxiv.org/abs/2411.15114}{RE-Bench} type
  stuff}
\item
  {Skills evidenced by public outputs like GitHub repositories}
\item
  {Ability to communicate technical progress effectively}
\end{itemize}

\end{minipage} \\
\end{longtable}
\normalsize
\renewcommand{\arraystretch}{1.2}

\clearpage
\subsection{Complete Company Evaluation}

\scriptsize
\setlength{\tabcolsep}{2.5pt}
\renewcommand{\arraystretch}{0.95}
\rowcolors{2}{rowshade}{white}
\begin{longtable}[]{@{}
  >{\raggedright\arraybackslash}p{0.85in}
  >{\centering\arraybackslash}p{0.30in}
  >{\raggedright\arraybackslash}p{0.55in}
  >{\raggedleft\arraybackslash}p{0.45in}
  >{\raggedright\arraybackslash}p{1.10in}
  >{\raggedleft\arraybackslash}p{0.75in}
  >{\raggedleft\arraybackslash}p{0.55in}
  >{\raggedright\arraybackslash}p{0.50in}
@{}}
\caption{Complete Company Evaluation: ML Research Talent estimates for all 403 identified organizations}\label{tab:company-eval}\\
\toprule
\textbf{Company} & \textbf{Year} & \textbf{Country} & \textbf{Staff} & \textbf{Individual Estimates} & \textbf{ML q50 (CI)} & \textbf{ML \%} & \textbf{Category} \\
\midrule
\endfirsthead

\multicolumn{8}{c}{\small\itshape Continued from previous page} \\
\toprule
\textbf{Company} & \textbf{Year} & \textbf{Country} & \textbf{Staff} & \textbf{Individual Estimates} & \textbf{ML q50 (CI)} & \textbf{ML \%} & \textbf{Category} \\
\midrule
\endhead

\midrule
\multicolumn{8}{r}{\small\itshape Continued on next page} \\
\endfoot

\bottomrule
\endlastfoot

\textbf{Tata Consultancy Services} & 1968 & India & 640 000 & {\tiny [305, 5400, 3200, 0, 0, 0]} & 87 (12 - 322) * & 0.01\% (0.00\% - 0.05\%) & Probable \\
\textbf{Accenture} & 1989 & Ireland & 550 000 & {\tiny [186, 3400, 3500, 0, 0, 0]} & 76 (13 - 280) * & 0.01\% (0.00\% - 0.05\%) & Probable \\
\textbf{Yandex} & 1997 & Russia & 9 900 & {\tiny [67, 521, 232, 142, 291, 528]} & 60 (41 - 90) & 0.61\% (0.41\% - 0.91\%) & Probable \\
\textbf{Deloitte} & 1845 & United Kingdom & 430 000 & {\tiny [171, 2200, 3700, 0, 0, 0]} & 54 (8 - 211) * & 0.01\% (0.00\% - 0.05\%) & Probable \\
\textbf{Cohere} & 2019 & Canada & 831 & {\tiny [45, 75, 187, 85, 112, 136]} & 45 (26 - 76) & 5.42\% (3.13\% - 9.15\%) & Probable \\
\textbf{Scale AI} & 2016 & United States & 5 800 & {\tiny [68, 221, 594, 82, 179, 326]} & 45 (29 - 75) & 0.78\% (0.50\% - 1.29\%) & Probable \\
\textbf{IBM} & 1911 & United States & 290 000 & {\tiny [443, 2300, 3900, -, -, -]} & 42 (5 - 143) * & 0.01\% (0.00\% - 0.05\%) & Probable \\
\textbf{Capgemini} & 1967 & France & 260 000 & {\tiny [103, 2100, 1500, -, -, -]} & 38 (5 - 131) * & 0.01\% (0.00\% - 0.05\%) & Probable \\
\end{longtable}
\normalsize
\renewcommand{\arraystretch}{1.2}
\rowcolors{1}{white}{white}
\setlength{\tabcolsep}{6pt}

\end{document}